\documentclass[10pt,onecolumn]{IEEEtran}
\usepackage{indentfirst}
\usepackage[dvips]{graphicx}
\usepackage{amsfonts}
\usepackage{multirow}
\usepackage{amsmath,amsthm,amssymb}
\usepackage{color,changepage}

\usepackage{algorithm}
\usepackage{algorithmic}
\usepackage{bbm}
\usepackage{verbatim}
\usepackage{makecell}
\usepackage{subfigure}
\usepackage{mathrsfs}
\usepackage{arydshln}
\usepackage{cases}

\usepackage{extarrows}
\usepackage{array}
\usepackage{bm}
\usepackage{pifont}
\usepackage[T1]{fontenc}
\usepackage{threeparttable}
\usepackage{booktabs}
\usepackage{caption}
\usepackage{mdwlist}
\allowdisplaybreaks[4]

\definecolor{newcolor}{rgb}{0.5,0,1}

\newcommand{\rightt}{\ding{52}}
\newcommand{\cmarkk}{\ding{56}}

\newcolumntype{I}{!{\vrule width 2pt}}

\newtheorem{Theorem}{Theorem}

\newtheorem{Lemma}{Lemma}

\theoremstyle{remark}
\newtheorem{Remark}{Remark}

\DeclareMathAlphabet{\mathpzc}{OT1}{pzc}{m}{it}

\definecolor{newcolor}{rgb}{0.5,0,1}

\begin{document}
\title{Symmetric Private Polynomial Computation From Lagrange Encoding}
\author{Jinbao Zhu, Qifa Yan, Xiaohu Tang, and Songze Li
\thanks{
J. Zhu, Q. Yan and X. Tang are with the Information Security and National Computing Grid Laboratory, Southwest Jiaotong University, Chengdu  611756, China (email: jinbaozhu@my.swjtu.edu.cn, qifayan@swjtu.edu.cn, xhutang@swjtu.edu.cn).
%

S. Li is with the IoT Thrust, The Hong Kong University of Science and Technology, China (e-mail: songzeli@ust.hk).
}
}
\maketitle

\begin{abstract}
The problem of $X$-secure $T$-colluding symmetric Private Polynomial Computation (PPC) from coded storage system with $B$ Byzantine and $U$ unresponsive servers is studied in this paper. Specifically, a dataset consisting of $M$ files is stored across $N$ distributed servers according to $(N,K+X)$ Maximum Distance Separable (MDS) codes such that any group of up to $X$ colluding servers can not learn anything about the data files. A user wishes to privately evaluate one out of a set of candidate polynomial functions over the $M$ files from the system, while guaranteeing that any $T$ colluding servers can not learn anything about the identity of the desired function and the user can not learn anything about the $M$ data files more than the desired polynomial function evaluations, in the presence of $B$ Byzantine servers that can send arbitrary responses maliciously to confuse the user and $U$ unresponsive servers that will not respond any information at all. A novel symmetric PPC scheme using Lagrange encoding is proposed. This scheme achieves a PPC rate of $1-\frac{G(K+X-1)+T+2B}{N-U}$ with secrecy rate $\frac{G(K+X-1)+T}{N-(G(K+X-1)+T+2B+U)}$ and finite field size $N+\max\{K,N-(G(K+X-1)+T+2B+U)\}$, where $G$ is the maximum degree over all the candidate polynomial functions. Moreover, to further measure the efficiency of PPC schemes, upload cost, query complexity, server computation complexity and decoding complexity required to implement the scheme are analyzed. Remarkably, the PPC setup studied in this paper generalizes all the previous MDS coded PPC setups and the degraded schemes strictly outperform the best known schemes in terms of (asymptotical) PPC rate, which is the main concern of the  PPC schemes.

\end{abstract}
\begin{IEEEkeywords}
Private information retrieval, symmetric private polynomial computation, Lagrange encoding, computation complexity.
\end{IEEEkeywords}

\section{Introduction}\label{Introduction}
With the rapid evolution of big data, machine learning and distributed computing, there arises  substantial concerns about protecting the computing privacy of a user from public servers. This problem is referred to as Private Computation (PC), which seeks efficient solutions for the user to compute a function of files  stored at distributed servers, without disclosing the identity of the desired function to the servers.
The PC problem was firstly introduced in \cite{PC2,PC1} and has attracted remarkable attention in the past few years within information-theoretic community \cite{PC4,PC3,PC5,Raviv PPC}.
In the classical PC setup, the user wishes to compute one out of any $P$ candidate functions over $M$ files from $N$ non-colluding servers, each of which stores all the $M$ files, while preventing any individual server
from obtaining information about which function is being computed. To this end, the user sends $N$ query strings, one to each server. 
After receiving the query, each server truthfully responds an answer string to the user based on the information it stores. Finally, the user is able to recover the desired function from the collected answer strings.

A trivial strategy
is to download all the files from the servers and then compute the desired function locally, or request severs to compute all the functions and then download all the evaluations, which incurs significant communication cost and therefore is highly impractical in practice. It was proved that the naive strategy is the only feasible solution in the sense of information-theoretic privacy if the files are stored at a single server \cite{Chor}. To alleviate this inefficiency, the information-theoretic PC with low communication cost can be achieved by replicating the files at multiple non-colluding servers \cite{Chor}.
In such systems, the most important measure of communication effectiveness is the PC rate, defined as the number of bits of the desired function that can be privately retrieved per  downloaded bit from all servers. The supremum of PC rates over all achievable schemes is referred to as its capacity. Indeed, private computation is a generalization of Private Information Retrieval (PIR) problem wherein the user wishes to privately retrieval one out of the $M$ files from the $N$ servers, while instead hiding the identity of the desired file from the servers,
see \cite{N_PIR5,Gasarch,Yekhanin} for PIR details.

In a recent influential work by Sun and Jafar \cite{PC1}, the exact capacity of classical Private Linear Computation (PLC) problem, where the user wants to privately compute a linear combination of the $M$ files, was characterized as $\big(1+\frac{1}{N}+\ldots+\frac{1}{N^{M-1}}\big)^{-1}$.
Soon afterwards, the problem of PLC over Maximum Distance Separable (MDS) coded storage (or MDS-PLC in short), where the files are distributed across the $N$ servers according to $(N,K)$ MDS codes, was considered by Obead \emph{et al.} in \cite{PC3} and its capacity was subsequently characterized  in \cite{MDS Obead} to be $\big(1+\frac{K}{N}+\ldots+\frac{K^{M-1}}{N^{M-1}}\big)^{-1}$.
Moreover,  they \cite{PC5} further constructed PLC schemes on arbitrary linear storage codes and showed that the capacity of MDS-PLC can be achieved for a large class of linear codes.

Particularly in \cite{PC5,PC4,Raviv PPC}, the problem of PC was focused on the setup that the candidate functions  are polynomials with maximum degree $G$ in $M$ variables (files) over a finite field $\mathbb{F}_q$, called Private Polynomial Computation (PPC). Very recently, Obead \emph{et al.} \cite{PC5} presented two novel non-colluding PPC schemes from systematic and nonsystematic Reed-Solomon coded servers for arbitrary number of candidate polynomial functions, also referred to as systematic MDS-PPC and nonsystematic MDS-PPC, respectively.
In \cite{PC4,Raviv PPC}, the $P$ candidate functions were restricted to be a finite-dimensional \emph{vector space (or sub-space)} of polynomials over $\mathbb{F}_q$.
Accordingly, Karpuk \cite{PC4} investigated PPC with $T$ colluding and systematically MDS coded servers (systematic MDS-TPPC), where any $T$ out of the $N$ servers can collude to deduce the identity of the interesting function, and proposed an MDS-TPPC scheme achieving the rate $\frac{\min\{N-(G(K-1)+T),K\}}{N}$ by generalizing the star-product PIR codes \cite{MDS-TPIR}.
Later in \cite{Raviv PPC}, the security setup was further generalized by Raviv and Karpuk to  the scenarios of $X$-secure data storage, $B$ Byzantine servers and $U$ unresponsive servers (U-B-MDS-XTPPC),
where the data security is guaranteed against up to $X$ colluding servers, while any group of up to $B$ servers return arbitrary responses maliciously to confuse the user 
and any $U$ disjoint servers do not respond any information at all. As a result, they  constructed an U-B-MDS-XTPPC scheme achieving the rate $\frac{N-(G(K+X-1)+T+2B+U)}{N-U}\cdot\frac{K}{G(K+X-1)+1}$ \cite{Raviv PPC} by leveraging ideas from Lagrange coded computation \cite{LCC,batch matrix} and successive decoding with interference cancellation strategy \cite{MDS-X-security,Tajeddine222}.


The problem of secure multi-party computation, first introduced by Yao in \cite{Yao}, focuses on jointly computing an arbitrary polynomial function of some private datasets distributed at the users (parties) under the constraint that each user must not gain any additional information about the datasets beyond the function  interested.
Naturally, it is also desirable to keep the data files private from the user more than the desired function in PC. For example, if one wishes to privately 
compute a feature function from massive medical datasets in medical big data, it is supposed to prevent he/she from learning anything about the medical records more than the desired function results, beside keeping the feature function private from the servers. This new constraint is called server-privacy and the corresponding problem is called symmetric PC.
To protect server-privacy, all the servers are allowed to share a common randomness that is independent of the files and unavailable to the user.
Consequently, secrecy rate is employed to be another metric to measure the effectiveness of symmetric PC schemes,
which is defined as the ratio of the amount of common randomness shared by the servers and the number of desired function evaluations.

In this paper, we consider the general problem of U-B-MDS-XTSPPC, i.e., symmetric private polynomial computation from $(N,K+X)$ MDS coded storage with $X$-secure data storage, $T$-colluding privacy, $B$ Byzantine servers and $U$ unresponsive servers, see Table \ref{Model:comparision} for the comparison with PPC schemes in previous setups, where vector space means that the candidate functions the user wishes to evaluate itself constitute a finite-dimensional vector space (or sub-space) of polynomials over $\mathbb{F}_q$, and arbitrary candidate functions mean that the polynomial functions the user wishes to evaluate are formed by arbitrary polynomials (it may not be a vector space of polynomials over $\mathbb{F}_q$).
Interestingly, we observed that the PPC schemes in \cite{PC4,Raviv PPC} can be straightly expanded to work on the general case of arbitrary candidate functions.
This is because the user can choose a vector space $\mathcal{P}$ of polynomials containing the original candidate functions, then the PPC schemes \cite{PC4,Raviv PPC} directly work over $\mathcal{P}$.
More precisely, when the user wishes to compute a desired polynomial function, to protect the privacy of the function, the user can first choose some random noises uniformly and independently from $\mathcal{P}$ and then employs the schemes \cite{PC4,Raviv PPC} to construct the queries sent to the servers, such that the quires are uniformly distributed on $\mathcal{P}$. It is easy to verify that the correctness and privacy can be guaranteed. Apparently, with larger dimension of the chosen polynomial space, the identity of the function being computed can be better protected, i.e., the possibility that the servers know which function is being computed is smaller.

\begin{table*}[!ht]
\centering
\begin{threeparttable}
  \begin{tabular}{|c|c|c|c|}
  \hline
   & MDS coded storage & $T$-Colluding Privacy & $X$-Security  \\ \hline
  Nonsystematic MDS-PPC Scheme \cite{PC5} & \rightt & \cmarkk & \cmarkk  \\ \hline
  Systematic MDS-PPC Scheme \cite{PC5} & \rightt & \cmarkk & \cmarkk \\ \hline
  Systematic MDS-TPPC Scheme \cite{PC4} & \rightt & \rightt & \cmarkk \\ \hline
  U-B-MDS-XTPPC Scheme \cite{Raviv PPC} & \rightt & \rightt & \rightt \\ \hline
  U-B-MDS-XTSPPC Scheme & \rightt & \rightt & \rightt \\ \hline
  \hline
  & Byzantine and Unresponsiveness & Server-Privacy & Candidate Polynomial Functions \\ \hline
  Nonsystematic MDS-PPC Scheme \cite{PC5} & \cmarkk & \cmarkk & Arbitrary  \\ \hline
  Systematic MDS-PPC Scheme \cite{PC5} & \cmarkk & \cmarkk & Arbitrary \\ \hline
  Systematic MDS-TPPC Scheme \cite{PC4} & \cmarkk & \cmarkk & Vector Space \\ \hline
  U-B-MDS-XTPPC Scheme \cite{Raviv PPC} & \rightt & \cmarkk & Vector Space \\ \hline
  U-B-MDS-XTSPPC Scheme & \rightt & \rightt & Arbitrary \\ \hline
  \end{tabular}
  \caption{\textit{Comparison for PPC schemes in previous setups.  The U-B-MDS-XTSPPC setup studied in this paper generalizes the previous setups of MDS-PPC \cite{PC5}, MDS-TPPC \cite{PC4}, and U-B-MDS-XTPPC \cite{Raviv PPC}.}} \label{Model:comparision}
\end{threeparttable}
\end{table*}

In PPC, the computation complexities, consisting of generating queries at user,  computing answers at severs and decoding  at user, should be considered to further measure the efficiency of PPC schemes.
The upload cost (the total length of query strings) and the size of finite field $\mathbb{F}_q$ operated by PC schemes are other  two important practical  design factors.
For these reasons, the objective of this paper is to design U-B-MDS-XTSPPC schemes with PPC rate as high as possible, while keeping secrecy rate, upload cost, finite field size, query complexity, server computation complexity and decoding complexity as small/low as possible.

As a result, we propose a novel U-B-MDS-XTSPPC scheme using Lagrange encoding \cite{LCC}. The scheme operates over the general case of arbitrary candidate polynomial functions, achieving the PPC rate $1-\frac{G(K+X-1)+T+2B}{N-U}$, secrecy rate $\frac{G(K+X-1)+T}{N-(G(K+X-1)+T+2B+U)}$, and finite field size $N+\max\{K,N-(G(K+X-1)+T+2B+U)\}$. Moreover, upload cost, query complexity, server computation complexity and decoding complexity required to implement the proposed scheme are also analysed.
Notably, with respect to the PPC rate, which is the main measure of PPC problem, our degraded schemes are strictly superior to the previous best known schemes for U-B-MDS-XTPPC \cite{Raviv PPC}, MDS-TPPC \cite{PC4}, and asymptotic MDS-PPC (i.e., the number of files $M\rightarrow\infty$) \cite{PC5}, see Section \ref{sec:comparison} for details.
In addition, the U-B-MDS-XTPPC scheme in \cite{Raviv PPC} and PIR schemes in \cite{Tajeddine222,MDS-X-security} require the queries, answers and decoding to happen over multi-rounds, and \emph{successive decoding} with interference cancellation strategy is employed by the user, i.e., the user will cancel the interference from the decoded information of previous rounds.
However, our scheme can be carried out \emph{independently and concurrently}, which improves the efficiency of retrieving desired information.

The rest of this paper is organized as follows. In Section \ref{system model}, the problem of U-B-MDS-XTSPPC is formally formulated.
In Section \ref{PPC scheme1}, the proposed U-B-MDS-XTSPPC scheme with Lagrange encoding is constructed for arbitrary candidate functions.
In Section \ref{proof:privacy2}, the feasibility of the scheme and its performance are analysed.
In Section \ref{sec:comparison}, the proposed scheme is compared with known results for the degraded problems of MDS-PPC, MDS-TPPC, and U-B-MDS-XTPPC.
Finally, the paper is concluded  in Section \ref{conclusion}.

The following notations are used throughout this paper.
\begin{itemize}
    \item Let boldface capital and lower-case letters represent matrices and vectors, respectively, e.g., $\mathbf{W}$ and $\mathbf{q}$;
  \item For any positive integers $m,n$ such that $m\leq n$, $[n]$ and $[m:n]$ denote the set $\{1,2,\ldots,n\}$ and $\{m,m+1,\ldots,n\}$, respectively;
  \item Define $A_{\Gamma}$ as $\{A_{\gamma_1},\ldots,A_{\gamma_{m}}\}$ for any index set $\Gamma=\{\gamma_1,\ldots,\gamma_{m}\}\subseteq[n]$;
  \item For a finite set $\mathcal{X}$, $|\mathcal{X}|$ denotes its cardinality.
\end{itemize}

\section{System Model}\label{system model}

Consider a dataset that comprises $M$ independent files,  $\mathbf{W}^{(1)},\ldots,\mathbf{W}^{(M)}$, and is stored at a distributed system with $N$ servers.
We assume that each file is divided into $K$ blocks, and each block is divided into $L$ stripes.\footnote{We divide the file into $K$ blocks because each file is stored at the distributed storage system according to MDS codes and $K$ is a \emph{fixed system parameter} of the MDS codes.
Moreover, in order to improve the flexibility of scheme design, each block is further divided into $L$ stripes, such that the user can efficiently retrieve desired function evaluations from server answers.
Notice that as is common in information theory, the file size is arbitrarily large and the coding scheme may freely choose the parameter $L$, i.e., $L$ is a \emph{free parameter} that needs to be carefully chosen to maximize the effectiveness of schemes.
Typically, such partitioning ideas have been widely applied in distributed storage system to reduce the repair bandwidth when repairing failed nodes from some surviving nodes \cite{A. G. Dimakis222,A. G. Dimakis}.}
WLOG, we represent the file $\mathbf{W}^{(m)}$  by a random matrix of dimension $L\times K$, with each entry chosen independently and uniformly  over the finite field $\mathbb{F}_q$ for some prime power $q$, i.e.,
\begin{IEEEeqnarray}{rCl}\label{file symbols}
\mathbf{W}^{(m)} =\left[
  \begin{array}{ccc}
    w^{(m)}_{1,1} & \ldots & w^{(m)}_{1,K}  \\
    \vdots & \ddots & \vdots \\
    w^{(m)}_{L,1} & \ldots &w^{(m)}_{L,K}\\
\end{array}
\right],\IEEEeqnarraynumspace \forall\,m\in[M].
\end{IEEEeqnarray}
The independence between all the files can be formalized as
\begin{IEEEeqnarray*}{rCl}
H( \mathbf{W}^{(1)},\ldots,\mathbf{W}^{(M)})&=&\sum_{m=1}^{M}H(\mathbf{W}^{(m)})=MLK, \label{model:file inden}
\end{IEEEeqnarray*}
where the entropy function $H(\cdot)$ is measured with logarithm $q$, and $LK$ is the number of symbols contained in each file.

The dataset is stored at the distributed system by using MDS codes over  $\mathbb{F}_q$ and kept secure 
 from any group of up to $X$ colluding servers.
In analogy to \cite{MDS-X-security,Raviv PPC}, security and MDS property are guaranteed by employing $(N,K+X)$ MDS codes.
Denote the information stored at server $n$ by $\mathbf{y}_n$ for any $n\in[N]$. Specifically, the storage system needs to satisfy
\begin{itemize}
  \item\textbf{MDS Property:} The dataset can be reconstructed by connecting to at least $K+X$ servers to tolerate up to $N-K-X$ server failures, i.e.,
    \begin{IEEEeqnarray}{c}\notag 
    H(\mathbf{W}^{(1)},\ldots,\mathbf{W}^{(M)}|\mathbf{y}_{\Gamma})=0,\quad\forall\,\Gamma\subseteq[N],|\Gamma|\geq K+X.
    \end{IEEEeqnarray}
    The storage at each server is constrained as $ML$, which is reduced by a factor of $\frac{1}{K}$ compared to repetition coding storage, i.e.,
    \begin{IEEEeqnarray}{rCl}
    H(\mathbf{y}_n)=ML, \quad\forall\,n\in[N].\notag
    \end{IEEEeqnarray}
  \item \textbf{$X$-Security:} Any $X$ servers remain oblivious perfectly to the dataset even if they collude, i.e.,
    \begin{IEEEeqnarray}{rCl}\label{X security}
    I(\mathbf{y}_{\mathcal{X}};\mathbf{W}^{(1)},\ldots,\mathbf{W}^{(M)})=0, \quad\forall\,\mathcal{X}\subseteq[N],|\mathcal{X}|=X.
    \end{IEEEeqnarray}
\end{itemize}
Obviously, the storage system degrades to the classical $(N,K)$ MDS coded setup when $X=0$.

Let $\phi^{(u)}(x_1,\ldots,x_M)\in\mathbb{F}_q[x_1,\ldots,x_M],u\in[P]$ be $P$ \emph{candidate multivariable polynomial functions} and $G$ be the maximum degree,
i.e.,
\begin{IEEEeqnarray}{rCl}\notag
G=\max\{\deg(\phi^{(u)}):u\in[P]\}.
\end{IEEEeqnarray}

In Private Polynomial Computation (PPC), a user privately selects a number $\theta\in[P]$  to evaluate  the   polynomial $\phi^{(\theta)}$ over the $M$ files $\mathbf{W}^{(1)},\ldots,\mathbf{W}^{(M)}$
from the $N$ servers, while keeping the index $\theta$ private from any colluding subset of up to $T$ out of the $N$ servers.
Here, the privacy of the user is restricted to $[P]$.
That is, each of servers knows the set of candidate polynomial functions $\{\phi^{(u)}\}_{u\in[P]}$, but any $T$ colluding servers can not learn any information about which function is being computed other than it being in $\{\phi^{(u)}\}_{u\in[P]}$.
Let $\mathbf{V}^{(\theta)}\triangleq\phi^{(\theta)}(\mathbf{W}^{(1)},\ldots,\mathbf{W}^{(M)})$ be the desired function evaluations of the user,
where $\mathbf{V}^{(\theta)}$ is an $L\times K$ random matrix of the form 
\begin{IEEEeqnarray}{rCl}\label{desired file}
\mathbf{V}^{(\theta)} =\left[
  \begin{array}{ccc}
    v^{(\theta)}_{1,1} & \ldots & v^{(\theta)}_{1,K}  \\
    \vdots & \ddots & \vdots \\
    v^{(\theta)}_{L,1} & \ldots &v^{(\theta)}_{L,K}\\
\end{array}
\right],\quad\forall\,\theta\in[P]
\end{IEEEeqnarray}
with
\begin{IEEEeqnarray}{rCl}\label{desired symbols}
v^{(\theta)}_{\ell,k}=\phi^{(\theta)}({w}^{(1)}_{\ell,k},\ldots,{w}^{(M)}_{\ell,k}),\IEEEeqnarraynumspace \forall\,\ell\in[L],k\in[K].
\end{IEEEeqnarray}

For this purpose, the user sends $S$ queries to each server, which accordingly responds the user with $S$ answers according to the information available. Consequently, the user is able to decode the desired evaluations from the answers of servers.
In addition, it is required that the user must not gain any information about the data files $\mathbf{W}^{(1)},\ldots,\mathbf{W}^{(M)}$ more than the desired evaluations, called \emph{symmetric} PPC.
This is guaranteed by the assumption that all the servers share a common randomness $\mathcal{F}$, which is independent of all the stored information $\mathbf{y}_{[N]}$ but unavailable to the user.

For convenience,  we divide the queries, answers and decoding into  $S$ \emph{rounds} or \emph{iterations}. During each round $s\in[S]$, we assume the presence of some servers $\mathcal{B}^{s}$ of size at most $B$ that pretend to send arbitrary  answers to confuse the user, known as \emph{Byzantine servers}, and another set of disjoint servers $\mathcal{U}^{s}$ of size at most $U$ that do not respond at all, known as \emph{unresponsive servers},  where the identities of the servers which are Byzantine and unresponsive  may change from round to round.  Note that the user has no priori knowledge of the identities of the Byzantine servers $\mathcal{B}^{s}$ and unresponsive servers $\mathcal{U}^{s}$, other than knowing  the values of $B$ and $U$.

Formally, an $X$-secure $T$-colluding Symmetric PPC scheme from MDS coded storage system with $B$ Byzantine and $U$ unresponsive servers, also referred to as U-B-MDS-XTSPPC scheme, is composed of the queries, answers and decoding of $S$ rounds, and each of rounds $s\in[S]$ is described as follows.
\begin{enumerate}
\item \emph{Query Phase:} The user generates $N$ queries $\mathbf{q}_{[N]}^{s}$ and sends $\mathbf{q}_{n}^{s}$ to server $n$ for all $n\in[N]$.
\item \emph{Answer Phase:} Upon receiving the query $\mathbf{q}_n^s$, a Byzantine server $n\in\mathcal{B}^{s}$ overwrites its answer maliciously and sends an arbitrary response $A_{n}^{s}$ to confuse the user, where  $\mathcal{B}^{s}\subseteq[N],|\mathcal{B}^{s}|\leq B$. An unresponsive server in $\mathcal{U}^{s}$ will not respond any information at all, where $\mathcal{U}^{s}\subseteq[N],\mathcal{U}^{s}\cap\mathcal{B}^{s}=\emptyset,|\mathcal{U}^{s}|\leq U$.
And the remaining servers in $[N]\backslash(\mathcal{B}^{s}\cup\mathcal{U}^{s})$, known as \emph{authentic servers}, will truthfully respond the answers, which are the determined functions of the received queries and the stored information, i.e.,
\begin{IEEEeqnarray}{c}\label{asnwers:123}
H(A_{n}^{s}|\mathbf{q}_{n}^{s},\mathbf{y}_n,\mathcal{F})=0,\quad \forall\,n\in[N]\backslash(\mathcal{B}^{s}\cup\mathcal{U}^{s}).
\end{IEEEeqnarray}
\item \emph{Decoding Phase:} The user decodes some interested data $\mathcal{V}^s$ from the information available to it in round $s$, i.e.,
\begin{IEEEeqnarray}{c}\notag
H(\mathcal{V}^s|\{A_{[N]\backslash\mathcal{U}^{s'}}^{s'},\mathbf{q}_{[N]}^{s'}\}_{s'\in[s]})=0.
\end{IEEEeqnarray}
\end{enumerate}

The following conditions must hold for an U-B-MDS-XTSPPC scheme.
\begin{itemize}
  \item\textbf{Correctness:} The desired function evaluations $\mathbf{V}^{(\theta)}$ must be obtained by converging the decoded data over the $S$ rounds, i.e.,
\begin{IEEEeqnarray}{c}\label{correct:}
H(\mathbf{V}^{(\theta)}|\{\mathcal{V}^s\}_{s\in[S]})=0,\quad\forall\, \theta\in[P].
\end{IEEEeqnarray}
\item\textbf{User-Privacy:} The desired function index $\theta$ must be hidden from all the queries sent to any $T$ colluding servers, i.e.,
\begin{IEEEeqnarray}{c}\label{Infor:priva cons}
I(\{\mathbf{q}_{\mathcal{T}}^{s}\}_{s\in[S]};\theta)=0,\quad\forall\, \mathcal{T}\subseteq[N],|\mathcal{T}|=T.
\end{IEEEeqnarray}
\item\textbf{Server-Privacy:} The user must not gain any additional information in regard to all the data files more than the desired polynomial function evaluations, i.e.,
\begin{IEEEeqnarray}{c}\label{Infor:priva cons2}
I(\{A_{[N]\backslash\mathcal{U}^{s}}^{s},\mathbf{q}_{[N]}^{s}\}_{s\in[S]};\mathbf{W}^{(1)},\ldots,\mathbf{W}^{(M)}|\mathbf{V}^{(\theta)})=0,\quad\forall\, \theta\in[P].
\end{IEEEeqnarray}
\end{itemize}


The performance of an U-B-MDS-XTSPPC scheme can be measured by the following five quantities:
\begin{enumerate}
  \item[1.] The PPC rate, which is the ratio of the number of desired function evaluations to the total downloaded symbols,
  defined as
  \begin{IEEEeqnarray}{c}\label{def:PPC rate}
  R_p\triangleq\frac{LK}{D},
  \end{IEEEeqnarray}
  where $D=\sum_{s\in[S]}\sum_{n\in[N]\backslash\mathcal{U}^{s}}H(A_{n}^{s})$ is the average download cost from the responsive servers over all  rounds.
  \item[2.] The secrecy rate, which is defined as the ratio of the amount of common randomness shared by the servers and the number of desired function evaluations, i.e.,
  \begin{IEEEeqnarray}{c}\label{def:Secrecy rate}
  R_s\triangleq\frac{H(\mathcal{F})}{LK}.
  \end{IEEEeqnarray}
  \item[3.] The upload cost, which is the number of symbols  required to send the queries to the servers, i.e.,
  \begin{IEEEeqnarray*}{c}\label{upload cost}
  C_u\triangleq \sum\limits_{s\in[S]}\sum\limits_{n\in[N]}H(\mathbf{q}_{n}^{s}).
  \end{IEEEeqnarray*}
  \item[4.] The finite field size $q$, which ensures the achievability of the storage codes and the coded PPC scheme.
  \item[5.] The system complexity, which includes the complexities of queries, server computation and decoding.
  Define the query complexity $\mathcal{C}_{q}$ at the user as the order of the number of arithmetic operations required to generate all the queries $\{\mathbf{q}_{[N]}^{s}\}_{s\in[S]}$.
  Similarly, define the server computation complexity $\mathcal{C}_{s}$ to be the order of the number of arithmetic operations required to generate the response $\{A_n^{s}\}_{s\in[S]}$, maximized over $n\in[N]$.
  Finally, define the decoding complexity $\mathcal{C}_d$ at the user to be the order of the number of arithmetic operations required to decode the desired function evaluations $\mathbf{V}^{(\theta)}$ from the answers of responsive servers.
  \end{enumerate}

In principle, the PPC rate is preferred to be high, while the secrecy rate, upload cost, finite field size and system complexity are preferred to be small/low.



\begin{Remark}\label{remark:degrade}
Different from all the previous MDS-PPC works \cite{PC4,PC5,Raviv PPC}, we consider the  generic U-B-MDS-XTSPPC problem.
When server-privacy is not considered (i.e., the constrain \eqref{Infor:priva cons2} is removed and thus the common randomness $\mathcal{F}$ is not necessary and the secrecy rate $R_s$ can be set to be $0$),
our problem straightly degrades to the problems of MDS-PPC \cite{PC5} by setting $U=B=X=0$ and $T=1$, MDS-TPPC \cite{PC4} by setting $U=B=X=0$, and U-B-MDS-XTPPC \cite{Raviv PPC}. In general, U-B-MDS-XTSPPC is an integration and generalization of previous studies on PPC extensions.
\end{Remark}

For clarity, the parameters used in our U-B-MDS-XTSPPC system are listed in Table \ref{tab:parameters}.

\begin{table*}[htbp]
\extrarowheight=4pt
\centering
\caption{Parameters Used in U-B-MDS-XTSPPC}
  \begin{tabular}{|c|c||c|c|}
  \hline
  $N$ & number of servers & $M$ & number of files \\ 
  \hline
  $L$ & number of rows of each data file & $K$ & number of columns of each data file \\
  \hline
  $X$ & number of colluding data-curious servers & $T$ & number of colluding function-curious servers \\
  \hline
  $B$ & number of Byzantine servers & $U$ & number of unresponsive servers \\
  \hline
  $P$ & number of candidate polynomial functions & $G$ & maximum degree over candidate polynomial functions \\
  \hline
  $q$ & finite field size   & $\mathbf{V}^{(\theta)}$ & desired polynomial function evaluations  \\
  \hline
  $\mathcal{F}$ & common randomness across servers & $S$ & number of rounds \\
  \hline
  $R_p$ & PPC rate & $R_s$ & secrecy rate \\
  \hline
  $C_u$  & upload cost & $\mathcal{C}_q$ & query complexity at user \\ \hline
  $\mathcal{C}_s$  & server computation complexity & $\mathcal{C}_d$ & decoding complexity at user \\
  \hline
  \end{tabular}
  \label{tab:parameters}
\end{table*}


\section{U-B-MDS-XTSPPC Scheme Based on Lagrange Encoding}\label{PPC scheme1}
In this section, we present an U-B-MDS-XTSPPC scheme based on Lagrange encoding, which works for the general case of any $P$ candidate polynomial functions.

Before that, we first introduce three useful lemmas, which will be employed by the later U-B-MDS-XTSPPC scheme to preserve/resist $X$-security, Byzantine and unresponsiveness, and user-privacy.

\begin{Lemma}[\cite{Shamir}]\label{lemma:security}
Given any positive integers $N,K,X$ such that $N\geq K+X$, let $w_1,\ldots,w_K\in\mathbb{F}_q$ be $K$ secrets and $z_1,\ldots,z_X$ be $X$ random variables chosen independently and uniformly from $\mathbb{F}_q$. Let $\alpha_1,\ldots,\alpha_N$ be $N$ distinct numbers from $\mathbb{F}_q$.
Denote
\begin{IEEEeqnarray}{c}\notag
\varphi(\alpha)={w}_{1}h_1(\alpha)+\ldots+{w}_{K}h_K(\alpha)+{z}_{1}c_1(\alpha)+\ldots+{z}_{X}c_X(\alpha),
\end{IEEEeqnarray}
where $h_1(\alpha),\ldots,h_K(\alpha),c_1(\alpha),\ldots,c_X(\alpha)$ are the deterministic function of $\alpha$.
If the matrix
\begin{IEEEeqnarray}{c}\notag
\mathbf{C}=
\left[
  \begin{array}{cccc}
    c_1(\alpha_{n_1}) & c_2(\alpha_{n_1}) & \ldots & c_X(\alpha_{n_1}) \\
    c_1(\alpha_{n_2}) & c_2(\alpha_{n_2}) & \ldots & c_X(\alpha_{n_2}) \\
    \vdots & \vdots & \ddots & \vdots \\
    c_1(\alpha_{n_X}) & c_2(\alpha_{n_X}) & \ldots & c_X(\alpha_{n_X}) \\
  \end{array}
\right]_{X\times X}
\end{IEEEeqnarray}
is non-singular over $\mathbb{F}_q$ for any $\mathcal{X}=\{n_1,\ldots,n_X\}\subseteq[N]$ with $|\mathcal{X}|=X$, then the $X$ values $\{{\varphi}(\alpha_{n_1}),\ldots,{\varphi}(\alpha_{n_X})\}$  can not reveal any information about the $K$ secrets ${w}_{1},\ldots,w_{K}$, i.e.,
\begin{IEEEeqnarray}{c}\notag
I({\varphi}(\alpha_{n_1}),\ldots,{\varphi}(\alpha_{n_X});{w}_{1},\ldots,{w}_{K})=0,\quad\forall\,\mathcal{X}=\{n_1,\ldots,n_X\}\subseteq[N],|\mathcal{X}|=X.
\end{IEEEeqnarray}
\end{Lemma}

\begin{Lemma}[\cite{Lin}]\label{property:codes}
An $(n,k)$ maximum distance separable code with dimension $k$ and length $n$ is capable of resisting $b$ Byzantine errors and $u$ unresponsive errors if $d_{\min}=n-k+1\geq 2b+u+1$.
\end{Lemma}

\begin{Lemma}[Generalized Cauchy Matrix \cite{Lin}]\label{g-cauchy matrix}
Let $\alpha_1,\ldots,\alpha_k$ and $\beta_1,\ldots,\beta_k$ be the elements from $\mathbb{F}_q$ such that $\alpha_i\neq\alpha_j,\beta_i\neq\beta_j$ for any $i,j\in[k]$ with $i\neq j$, and $v_1,\ldots,v_k$ be $k$ nonzero elements from $\mathbb{F}_q$. Denote by $f_i(\alpha)$ a polynomial of degree $k-1$
\begin{IEEEeqnarray}{c}\notag
f_i(\alpha)=\prod\limits_{j\in[k]\backslash\{i\}}\frac{\alpha-\beta_j}{\beta_i-\beta_j},\quad\forall\,i\in[k].
\end{IEEEeqnarray}
Then the following generalized Cauchy matrix $\mathbf{F}_c$ is invertible over $\mathbb{F}_q$.
\begin{IEEEeqnarray}{c}\notag
\mathbf{F}_c=
\left[
  \begin{array}{cccc}
    v_1 f_{1}(\alpha_1) & v_1 f_{2}(\alpha_1)& \ldots & v_1 f_{k}(\alpha_1)  \\
    v_2 f_{1}(\alpha_2) & v_2 f_{2}(\alpha_2)& \ldots & v_2 f_{k}(\alpha_2)  \\
    \vdots & \vdots & \ddots & \vdots \\
    v_k f_{1}(\alpha_k) & v_k f_{2}(\alpha_k)& \ldots & v_k f_{k}(\alpha_k)  \\
\end{array}
\right].
\end{IEEEeqnarray}
\end{Lemma}

In each round, our scheme just allows each server to respond one symbol except for the unresponsive servers.
In order to efficiently resist the Byzantine errors and unresponsive errors such that the user can maximally retrieve the desired function evaluations from the answers during each round, it is desirable to enable the responses of all the servers to constitute an MDS codeword because it has maximum minimum Hamming distance.
Intuitively, among the server responses of $N$ dimensions in each round, our scheme exploits $T$ dimensions to preserve user-privacy and $G(K+X-1)$ dimensions to completely eliminate the uncertainty incurred by evaluating the polynomial function of degree $G$ at the $(N,K+X)$ MDS coded data.
In addition, $2B+U$ dimensions are used to correct the $B$ Byzantine errors and $U$ unresponsive errors.
Accordingly, the remaining $N-(G(K+X-1)+T+2B+U)$ dimensions are left for us to retrieve desired function evaluations.

Given any PPC scheme, let $E$ denote the number of desired function evaluations that the user can privately retrieve in each round of the scheme. In our PPC scheme, the server responses of the remaining $N-(G(K+X-1)+T+2B+U)$ dimensions in each round are completely exploited to retrieve desired function evaluations, i.e., our scheme sets
\begin{IEEEeqnarray}{c}\label{para:def}
E= N-(G(K+X-1)+T+2B+U)
\end{IEEEeqnarray}
with $N>G(K+X-1)+T+2B+U$.
Recall from \eqref{desired file} that the user needs to compute $LK$ polynomial function evaluations.
The parameter $L$ and number of rounds $S$ should satisfy
\begin{IEEEeqnarray}{c}\label{paramters}
ES=LK.
\end{IEEEeqnarray}
Here, we choose the smallest integers satisfying \eqref{paramters}, i.e.,
\begin{IEEEeqnarray}{c}\label{chose:parameters}
L=\frac{E}{\Delta},\quad S=\frac{K}{\Delta},
\end{IEEEeqnarray}
where $\Delta\triangleq\gcd(K,E)$.

\subsection{Public Elements}\label{Achievable Parameters}
To construct U-B-MDS-XTSPPC scheme, we first need to generate elements $\{\beta_{\ell,k}:\ell\in[L],k\in[K+X]\}$ and $\{\alpha_1,\ldots,\alpha_N\}$ from $\mathbb{F}_q$, which will be publicized to the user and severs in advance. 

Denote $\{\beta_{\ell,k}:\ell\in[L],k\in[K+X]\}\subseteq\mathbb{F}_q$ by a matrix $\boldsymbol{\beta}$ of dimension $L\times(K+X)$, i.e.,
\begin{IEEEeqnarray*}{rCl}
\boldsymbol{\beta}\triangleq\left[
  \begin{array}{ccc;{2pt/2pt}ccc}
    \beta_{1,1} & \ldots & \beta_{1,K} & \beta_{1,K+1} & \ldots & \beta_{1,K+X}  \\
    \vdots & \ddots & \vdots & \vdots & \ddots & \vdots \\
    \beta_{ L ,1} & \ldots &\beta_{ L ,K} & \beta_{ L ,K+1} & \ldots &\beta_{ L ,K+X} \\
\end{array}
\right].
\end{IEEEeqnarray*}
Throughout this paper, let $\{\beta_{\ell,k},\alpha_n:\ell\in[L],k\in[K+X],n\in[N]\}$  satisfy
\begin{enumerate}
  \item[P1.] The entries in each row of the matrix $\boldsymbol{\beta}$ are pairwise distinct, i.e., for each given $\ell\in[L]$, $\beta_{\ell,j}\neq\beta_{\ell,k}$ for all $j,k\in[K+X]$ with $j\neq k$;
  \item[P2.] For any given $s\in[S]$, all the entries in columns $[(s-1)\Delta+1:s\Delta]$ of the matrix $\boldsymbol{\beta}$ are pairwise distinct, i.e., $\beta_{\ell,k}\neq\beta_{i,j}$ for any $(\ell,k)\neq(i,j)$  such that $\ell,i\in[L]$ and $k,j\in[(s-1)\Delta+1:s\Delta]$;
  \item[P3.] The elements $\alpha_1,\ldots,\alpha_N$ are distinct, i.e., $\alpha_i\neq\alpha_j$ for all $i,j\in[N]$ with $i\neq j$;
  \item[P4.] The elements $\alpha_1,\ldots,\alpha_N$ are distinct from the ones in columns $[K]$ of the matrix $\boldsymbol{\beta}$, i.e., $\{\alpha_n:n\in[N]\}\cap\{\beta_{\ell,k}:\ell\in[L],k\in[K]\}=\emptyset$.
\end{enumerate}

The following lemma states the sufficient condition of the finite field for finding such elements, which will be proved in Appendix.  
\begin{Lemma}\label{theorem:tield size}
There must exist a group of elements $\{\beta_{\ell,k},\alpha_n:\ell\in[L],k\in[K+X],n\in[N]\}\subseteq\mathbb{F}_q$ satisfying P1-P4 if $q\geq N+\max\{K, E\}$. 
\end{Lemma}


\subsection{Secure Lagrange Storage Codes}\label{LSC}
In this subsection, we describe the data encoding procedures, which use Lagrange interpolation polynomials to encode each row data of each file separately.
For any $m\in[M],\ell\in[L]$, let $z_{\ell,K+1}^{(m)},z_{\ell,K+2}^{(m)},\ldots,z_{\ell,K+X}^{(m)}$ be $X$ random variables distributed independently and uniformly on $\mathbb{F}_q$.  
Similar to \cite{Raviv PPC,LCC}, choose a polynomial $\varphi_\ell^{(m)}(\alpha)$ of degree at most $K+X-1$ for every $m\in[M],\ell\in[L]$ such that
\begin{IEEEeqnarray}{c}\label{encdoing}
\varphi_\ell^{(m)}(\beta_{\ell,k})=\left\{
\begin{array}{@{}ll}
w_{\ell,k}^{(m)},&\forall\, k\in[K]\\
z_{\ell,k}^{(m)},&\forall\, k\in[K+1:K+X]
\end{array}\right.,
\end{IEEEeqnarray}
where $w_{\ell,k}^{(m)}$ is the $\ell$-th element in the $k$-th column of data file $\mathbf{W}^{(m)}$ defined in \eqref{file symbols}.

By P1, the Lagrange interpolation rule and the degree restriction guarantee the existence and uniqueness of $\varphi_\ell^{(m)}(\alpha)$, which is expressed as
\begin{IEEEeqnarray}{c}\label{storage code}
\varphi_\ell^{(m)}(\alpha)=\sum\limits_{i=1}^{K}w_{\ell,i}^{(m)}\cdot\prod_{j\in[K+X]\backslash\{i\}}\frac{\alpha-\beta_{\ell,j}}{\beta_{\ell,i}-\beta_{\ell,j}}+
\sum\limits_{i=K+1}^{K+X}z_{\ell,i}^{(m)}\cdot\prod_{j\in[K+X]\backslash\{i\}}\frac{\alpha-\beta_{\ell,j}}{\beta_{\ell,i}-\beta_{\ell,j}}.
\end{IEEEeqnarray}
Then the evaluations of $\varphi_\ell^{(m)}(\alpha)$ $(m\in[M],\ell\in[ L ])$ at point $\alpha=\alpha_n$ are stored at the $n$-th server for any $n\in[N]$, i.e.,
\begin{IEEEeqnarray}{c}\label{stored data}
\mathbf{y}_n=\left(\varphi_1^{(1)}(\alpha_n),\ldots,\varphi_{1}^{(M)}(\alpha_n),\ldots,\varphi_{ L }^{(1)}(\alpha_n),\ldots,\varphi_{ L }^{(M)}(\alpha_n)\right).
\end{IEEEeqnarray}
Notice that, such Lagrange encoding is equivalent to the $(N,K+X)$ Reed-Solomon (RS) code \cite{Raviv PPC} with a class of specific basis polynomials $\sigma_{\ell,1}(z),\sigma_{\ell,2}(z),\ldots,\sigma_{\ell,K+X}(z)$ for any $\ell\in[ L ]$, where
\begin{IEEEeqnarray}{c}\notag
\sigma_{\ell,i}(z)=\prod_{j\in[K+X]\backslash\{i\}}\frac{\alpha-\beta_{\ell,j}}{\beta_{\ell,i}-\beta_{\ell,j}},\quad\forall\,i\in[K+X].
\end{IEEEeqnarray}
Hence, $\big(\varphi_\ell^{(m)}(\alpha_1),\ldots,\varphi_\ell^{(m)}(\alpha_N)\big)$ is an $(N,K+X)$ RS codeword over $\mathbb{F}_q$ for any $m\in[M],\ell\in[ L ]$ and the storage encoding has the property of $(N,K+X)$ MDS.

\begin{Remark}\label{remark:22}
In essence, to ensure data security and $(N,K+X)$ MDS property, the secure storage codes in our PPC scheme and \cite{Raviv PPC} employ Lagrange interpolation polynomials to encode $K$ data and $X$ random noises like \cite{LCC}. However, the storage codes in \cite{Raviv PPC} just use a group of coding parameters $(\beta_{1},\beta_{2},\ldots,\beta_{K+X})$ to encode each row data of all data files, which will cause the user not being able to distinguish all the desired function evaluations. For this purpose, our storage codes \eqref{encdoing} use $L$ groups of distinct coding parameters $\{(\beta_{\ell,1},\beta_{\ell,2},\ldots,\beta_{\ell,K+X}):\ell\in[L]\}$ satisfying P1-P4 to encode $L$ row data of each data file, respectively.
\end{Remark}

\subsection{Construction of U-B-MDS-XTSPPC Scheme}\label{s-PPC scheme}

Recall that the user wishes to privately compute the function evaluations $\mathbf{V}^{(\theta)}$ in \eqref{desired file}, where $\theta\in[P]$. To this end, the queries, answers and decoding of $S$ rounds will be operated as follows.

Before that, let $\mathcal{P}$ be the vector space spanned by the candidate polynomial functions $\{\phi^{(u)}\}_{u\in[P]}$ defined over $\mathbb{F}_q$, i.e.,
\begin{IEEEeqnarray}{rCl}
\mathcal{P}\triangleq\bigg\{\sum\limits_{u\in[P]}c_u\cdot\phi^{(u)}:c_1,c_2,\ldots,c_{P}\in\mathbb{F}_q\bigg\},
\end{IEEEeqnarray}
which will be used for protecting the user privacy.



During each round $s\in[S]$, the user independently and uniformly generates $LT$ random polynomial functions $\{\phi_{i,t}^{s}\}_{i\in[L],t\in[T]}$ from the polynomial space $\mathcal{P}$. Note that both $\phi^{(\theta)}$ and $\{\phi_{i,t}^{s}\}_{i\in[L],t\in[T]}$ are polynomials with $M$ variables $x_1,\ldots,x_M$.
Then, for each $i\in[L]$, construct the query polynomial $\rho_{i}^{s}(x_1,\ldots,x_M,\alpha)$ of degree $E+T-1=L\Delta+T-1$ in variable $\alpha$ such that
\begin{IEEEeqnarray}{rCll}
\rho_i^{s}(x_1,\ldots,x_M,\beta_{\ell,k})&=&
\left\{\begin{array}{@{}ll}
\phi^{(\theta)}(x_1,\ldots,x_M), &\mathrm{if}\,\,\ell=i\\
0, &  \mathrm{otherwise}
\end{array}
\right., &\quad\forall\, \ell\in[L],k\in[(s-1)\Delta+1:s\Delta],\label{scheme:2} \\
\rho_i^{s}(x_1,\ldots,x_M,\alpha_{t})&=&\phi_{i,t}^{s}(x_1,\ldots,x_M),&\quad\forall\, t\in[T]. \label{scheme:2222}
\end{IEEEeqnarray}
By P2-P4, the $L\Delta+T$ elements $\{\beta_{\ell,k}:\ell\in[L],k\in[(s-1)\Delta+1:s\Delta]\}\cup\{\alpha_{t}:t\in[T]\}$ are pairwise distinct for any $s\in[S]$. Thus, the polynomial $\rho_{i}^{s}(x_1,\ldots,x_M,\alpha)$ can be accurately written as
\begin{IEEEeqnarray}{rCl}
\rho_i^{s}(x_1,\ldots,x_M,\alpha)&=&\sum\limits_{l\in[T]}\phi_{i,l}^{s}(x_1,\ldots,x_M)\cdot\left(\prod\limits_{j\in[L],r\in[(s-1)\Delta+1:s\Delta]}\frac{\alpha-\beta_{j,r}}{\alpha_{l}-\beta_{j,r}}\right)\left(\prod\limits_{v\in[T]\backslash\{l\}}\frac{\alpha-\alpha_v}{\alpha_{l}-\alpha_v}\right) \notag\\
&&+\sum\limits_{l\in[(s-1)\Delta+1:s\Delta]}\phi^{(\theta)}(x_1,\ldots,x_M)\cdot\left(\prod\limits_{\substack{j\in[L],r\in[(s-1)\Delta+1:s\Delta]\\(j,r)\neq(i,l)}}\frac{\alpha-\beta_{j,r}}{\beta_{i,l}-\beta_{j,r}}\right)
\left(\prod\limits_{v\in[T]}\frac{\alpha-\alpha_v}{\beta_{i,l}-\alpha_v}\right).\label{query:2}
\end{IEEEeqnarray}


Then, the query sent to server $n$ is given by
\begin{IEEEeqnarray}{c}\label{query:22}
\mathbf{q}_n^{s}=\left( \rho_{1}^{s}(x_1,\ldots,x_M,\alpha_n),\rho_{2}^{s}(x_1,\ldots,x_M,\alpha_n),\ldots,\rho_{L}^{s}(x_1,\ldots,x_M,\alpha_n) \right),\quad\forall\,n\in[N],
\end{IEEEeqnarray}
where $\rho_{i}^{s}(x_1,\ldots,x_M,\alpha_n)$ is the evaluation of polynomial $\rho_{i}^{s}(x_1,\ldots,x_M,\alpha)$ at $\alpha=\alpha_n$ for any $i\in[L]$.
Accordingly, $\rho_{i}^{s}(x_1,\ldots,x_M,\alpha_n)$ is a linear combination of the polynomial functions $\phi^{(\theta)},\{\phi_{i,l}^{s}\}_{l\in[T]}$. Thus, $\rho_{i}^{s}(x_1,\ldots,x_M,\alpha_n)$ is also a polynomial belonging to the set $\mathcal{P}$ in $M$ variables because the polynomial function set $\mathcal{P}$ is a vector space over $\mathbb{F}_q$. 

Let
\begin{IEEEeqnarray}{c}\label{common:random}
\mathcal{F}^{s}=\{z_{j}^{s}:j\in[G(K+X-1)+T]\}
\end{IEEEeqnarray}
be $G(K+X-1)+T$ random variables distributed independently and uniformly over $\mathbb{F}_q$, which are shared by all the servers but unknown to the user. Define an interpolation polynomial $\psi^s(\alpha)$ of degree $E+G(K+X-1)+T-1$ such that
\begin{IEEEeqnarray}{rCll}
\psi^s(\beta_{\ell,k})&=&0,&\quad\forall\, \ell\in[L],k\in[(s-1)\Delta+1:s\Delta], \label{symmetric:1}\\
\psi^s(\alpha_{j})&=&z_{j}^{s},&\quad\forall\, j\in[G(K+X-1)+T]. \label{symmetric:122}
\end{IEEEeqnarray}
Note from P2-P4 again that $\{\beta_{\ell,k}:\ell\in[L],k\in[(s-1)\Delta+1:s\Delta]\}\cup\{\alpha_j:j\in[G(K+X-1)+T]\}$ are $E+G(K+X-1)+T$ distinct elements from $\mathbb{F}_q$ due to $G(K+X-1)+T<N$. Thus, the polynomial $\psi^s(\alpha)$ is the form of
\begin{IEEEeqnarray}{c}\label{symmetric:answer}
\psi^s(\alpha)\triangleq \sum\limits_{l\in[G(K+X-1)+T]}z_{l}^{s}\cdot \left(\prod\limits_{j\in[L],r\in[(s-1)\Delta+1:s\Delta]}\frac{\alpha-\beta_{j,r}}{\alpha_{l}-\beta_{j,r}}\right)\left(
\prod\limits_{v\in[G(K+X-1)+T]\backslash\{l\}}\frac{\alpha-\alpha_{v}}{\alpha_{l}-\alpha_{v}}\right).
\end{IEEEeqnarray}

Then, server $n$ computes a response by evaluating the query polynomials \eqref{query:22} received from the user at its stored data \eqref{stored data}, and then takes the sum of these evaluation results and another evaluation of $\psi^s(\alpha)$ at $\alpha=\alpha_n$, i.e.,
\begin{IEEEeqnarray}{rCl}
A^s_n&=&\sum\limits_{i=1}^{L}\rho_{i}^{s}(\varphi_i^{(1)}(\alpha_n),\ldots,\varphi_{i}^{(M)}(\alpha_n),\alpha_n)+\psi^s(\alpha_n). \label{answer:22}
\end{IEEEeqnarray}
Note that there are at most $B$ Byzantine servers, each of which instead generates an arbitrary element from $\mathbb{F}_q$ to confuse the user.
Meanwhile, there are at most $U$ unresponsive servers that will not respond any information at all.

Denote the answer polynomial $\zeta^s(\alpha)$ by
\begin{IEEEeqnarray}{rCl}
\zeta^s(\alpha)&=&\sum\limits_{i=1}^{ L }\rho_{i}^{s}(\varphi_i^{(1)}(\alpha),\ldots,\varphi_{i}^{(M)}(\alpha),\alpha)+\psi^s(\alpha)\label{polynomial:scheme2}\\
&=&\sum\limits_{i=1}^{L}\sum\limits_{l\in[T]}\phi_{i,l}^{s}(\varphi_i^{(1)}(\alpha),\ldots,\varphi_{i}^{(M)}(\alpha))\cdot\left(\prod\limits_{j\in[L],r\in[(s-1)\Delta+1:s\Delta]}\frac{\alpha-\beta_{j,r}}{\alpha_{l}-\beta_{j,r}}\right)\left(\prod\limits_{v\in[T]\backslash\{l\}}\frac{\alpha-\alpha_v}{\alpha_{l}-\alpha_v}\right)\notag\\
&&+\sum\limits_{i=1}^{L}\sum\limits_{l\in[(s-1)\Delta+1:s\Delta]}\phi^{(\theta)}(\varphi_i^{(1)}(\alpha),\ldots,\varphi_{i}^{(M)}(\alpha))\cdot\left(\prod\limits_{\substack{j\in[L],r\in[(s-1)\Delta+1:s\Delta]\\(j,r)\neq(i,l)}}\frac{\alpha-\beta_{j,r}}{\beta_{i,l}-\beta_{j,r}}\right)
\left(\prod\limits_{v\in[T]}\frac{\alpha-\alpha_v}{\beta_{i,l}-\alpha_v}\right)+\psi^s(\alpha).\notag
\end{IEEEeqnarray}
Obviously, the answer $A_n^s$ is equivalent to evaluating $\zeta^s(\alpha)$ at $\alpha=\alpha_n$ for any authentic server $n\in[N]\backslash(\mathcal{B}^{s}\cup\mathcal{U}^{s})$.
Since $\phi^{s}_{i,l},\phi^{(\theta)}$ are the polynomials in $M$ variables with degree at most $G$ for any $i\in[L],l\in[T],\theta\in[P]$ and the degree of polynomial $\varphi_{i}^{(m)}(\alpha)$ is $K+X-1$ for any $m\in[M],i\in[L]$ by \eqref{storage code},
the composite polynomials $\phi_{i,l}^{s}(\varphi_i^{(1)}(\alpha),\ldots,\varphi_{i}^{(M)}(\alpha))$ and $\phi^{(\theta)}(\varphi_i^{(1)}(\alpha),\ldots,\varphi_{i}^{(M)}(\alpha))$ have degree at most $G(K+X-1)$.
Thus, $\zeta^s(\alpha)$ can be viewed as a polynomial of single variable $\alpha$ with degree $G(K+X-1)+E+T-1$.
Recall from P3 that $\{\alpha_n\}_{n\in[N]}$ are distinct elements from $\mathbb{F}_q$.
So, $(\zeta^s(\alpha_1),\ldots,\zeta^s(\alpha_N))$ forms an $(N,G(K+X-1)+E+T)$ RS codeword, which provides robustness against $B$ random errors and $U$ erasure errors  at the same time by \eqref{para:def} and Lemma \ref{property:codes}.
Then, the user can decode the polynomial $\zeta^s(\alpha)$ from the answers $(A_1^s,\ldots,A_N^s)=(\zeta^s(\alpha_1),\ldots,\zeta^s(\alpha_N))$ by using RS decoding algorithms \cite{Lin,Gao} even if there exists $B$ Byzantine servers and $U$ unresponsive servers.

By \eqref{polynomial:scheme2}, for any $\ell\in[L]$ and $k\in[(s-1)\Delta+1:s\Delta]$, the user evaluates $\zeta^s(\alpha)$ at $\alpha=\beta_{\ell,k}$ to obtain
\begin{IEEEeqnarray}{rCl}
\zeta^s(\beta_{\ell,k})&=&\sum\limits_{i=1}^{L}\rho_{i}^{s}(\varphi_i^{(1)}(\beta_{\ell,k}),\ldots,\varphi_{i}^{(M)}(\beta_{\ell,k}),\beta_{\ell,k})+\psi^s(\beta_{\ell,k}) \label{evaluating:12345}\\
&\overset{(a)}{=}&\phi^{(\theta)}(\varphi_\ell^{(1)}(\beta_{\ell,k}),\ldots,\varphi_{\ell}^{(M)}(\beta_{\ell,k}))+\psi^s(\beta_{\ell,k}) \\
&\overset{(b)}{=}&\phi^{(\theta)}(w_{\ell,k}^{(1)},\ldots,w_{\ell,k}^{(M)}) \\
&\overset{(c)}{=}&v^{(\theta)}_{\ell,k}, \label{evaluating:1}
\end{IEEEeqnarray}
where $(a)$ is due to $\rho_i^{s}(x_1,\ldots,x_M,\beta_{\ell,k})=\phi^{(\theta)}(x_1,\ldots,x_M)$ if $i=\ell$ and $\rho_i^{s}(x_1,\ldots,x_M,\beta_{\ell,k})=0$ otherwise for any $\ell\in[L],k\in[(s-1)\Delta+1:s\Delta]$ by \eqref{scheme:2};
$(b)$ follows by \eqref{encdoing} and \eqref{symmetric:1}; $(c)$ follows from \eqref{desired symbols}.

Therefore,  in round $s$, the desired evaluations in columns $[(s-1)\Delta+1:s\Delta]$ of $\mathbf{V}^{(\theta)}$ \eqref{desired file} can be obtained by evaluating $\zeta^s(\alpha)$ at $\beta_{\ell,k}$ for all $\ell\in[L],k\in[(s-1)\Delta+1:s\Delta]$, i.e.,
\begin{IEEEeqnarray}{c}\label{decoding each round}
\mathcal{V}^{s}=\{v^{(\theta)}_{\ell,k}:\ell\in[L],k\in[(s-1)\Delta+1:s\Delta]\}.
\end{IEEEeqnarray}
As a result, the user can decode $\mathbf{V}^{(\theta)}$ correctly after traversing $s\in[S]$, where $S=K/\Delta$ by \eqref{chose:parameters}.

\begin{Remark}
Evidently, the U-B-MDS-XTSPPC scheme allows the decoding of $S$ rounds to be carried out \emph{independently and concurrently}, which are exceedingly efficient for retrieving desired function evaluations.
\end{Remark}

\subsection{Illustrative Example for U-B-MDS-XTSPPC Scheme}\label{example:scheme}
In this subsection, we present an explicit example to illustrate the main ideas of the proposed U-B-MDS-XTSPPC scheme for the parameters $N=21,K=4,X=2,G=2,M=2,T=2,B=1,U=1$, where $E=6,\Delta=2,L=3$ and $S=2$.

\subsubsection*{Lagrange Data Encoding}The data encoding operates as follows. Let $\{\beta_{\ell,k},\alpha_n:\ell\in[3],k\in[6],n\in[21]\}\subseteq\mathbb{F}_q$ be a group of elements satisfying P1-P4. For any $m\in[2]$ and $\ell\in[3]$, choose $X=2$ random variables $z_{\ell,5}^{(m)},z_{\ell,6}^{(m)}$ independently and uniformly from $\mathbb{F}_q$ and design the Lagrange interpolation polynomial $\varphi_\ell^{(m)}(\alpha)$ of degree $K+X-1=5$ such that
\begin{IEEEeqnarray}{rClrClrCl}
\varphi_\ell^{(m)}(\beta_{\ell,1})&=&w_{\ell,1}^{(m)},\quad
\varphi_\ell^{(m)}(\beta_{\ell,2})&=&w_{\ell,2}^{(m)},\quad
\varphi_\ell^{(m)}(\beta_{\ell,3})&=&w_{\ell,3}^{(m)},\label{example:storage}\\
\varphi_\ell^{(m)}(\beta_{\ell,4})&=&w_{\ell,4}^{(m)},\quad
\varphi_\ell^{(m)}(\beta_{\ell,5})&=&z_{\ell,5}^{(m)},\quad
\varphi_\ell^{(m)}(\beta_{\ell,6})&=&z_{\ell,6}^{(m)}.
\end{IEEEeqnarray}
The data stored at server $n\in[21]$ is
\begin{IEEEeqnarray}{c}\label{example:21}
\mathbf{y}_n=\big(\varphi_1^{(1)}(\alpha_n),\varphi_1^{(2)}(\alpha_n),\varphi_2^{(1)}(\alpha_n),\varphi_2^{(2)}(\alpha_n),\varphi_3^{(1)}(\alpha_n),\varphi_3^{(2)}(\alpha_n)\big).
\end{IEEEeqnarray}
By \eqref{desired file}, the user wishes to compute the following polynomial evaluations from the system.
\begin{IEEEeqnarray}{c}
v^{(\theta)}_{\ell,k}=\phi^{(\theta)}({w}^{(1)}_{\ell,k},{w}^{(2)}_{\ell,k}),\quad  \forall\, \ell\in[3], k\in[4]. \label{example2:desired}
\end{IEEEeqnarray}

\subsubsection*{U-B-MDS-XTSPPC Scheme}  For this purpose, during round $s\in[2]$, the user independently and uniformly generates $LT=6$ polynomial functions $\phi_{1,1}^{s},\phi_{1,2}^{s},\phi_{2,1}^{s},\phi_{2,2}^{s},\phi_{3,1}^{s},\phi_{3,2}^{s}$ from the polynomial space spanned by the candidate polynomial functions $\{\phi^{(u)}\}_{u\in[P]}$ over $\mathbb{F}_q$.

For each $i\in[3]$, construct the query polynomial $\rho_i^{s}(x_1,x_2,\alpha)$ of degree $E+T-1=7$ in variable $\alpha$ as
\begin{IEEEeqnarray}{rCl}
\rho_i^{s}(x_1,x_2,\alpha)&=&\sum\limits_{l\in[2]}\phi_{i,l}^{s}(x_1,x_2)\cdot\left(\prod\limits_{j\in[3],r\in[2s-1:2s]}\frac{\alpha-\beta_{j,r}}{\alpha_{l}-\beta_{j,r}}\right)\left(\prod\limits_{v\in[2]\backslash\{l\}}\frac{\alpha-\alpha_v}{\alpha_{l}-\alpha_v}\right) \notag\\
&&+\sum\limits_{l\in[2s-1:2s]}\phi^{(\theta)}(x_1,x_2)\cdot\left(\prod\limits_{\substack{j\in[3],r\in[2s-1:2s]\\(j,r)\neq(i,l)}}\frac{\alpha-\beta_{j,r}}{\beta_{i,l}-\beta_{j,r}}\right)\left(\prod\limits_{v\in[2]}\frac{\alpha-\alpha_v}{\beta_{i,l}-\alpha_v}\right).\quad
\end{IEEEeqnarray}

Then, the query sent to server $n\in[21]$ is
\begin{IEEEeqnarray}{c}\notag
\mathbf{q}_n^{s}=\left( \rho_{1}^{s}(x_1,x_2,\alpha_n),\rho_{2}^{s}(x_1,x_2,\alpha_n),\rho_{3}^{s}(x_1,x_2,\alpha_n) \right),
\end{IEEEeqnarray}
where $\rho_{i}^{s}(x_1,x_2,\alpha_n)$ is the evaluation of polynomial $\rho_{i}^{s}(x_1,x_2,\alpha)$ at $\alpha=\alpha_n$ and thus can be viewed as a linear combination of the polynomial functions $\{\phi^{(\theta)},\phi_{i,1}^{s},\phi_{i,2}^{s}\}$ for any $i\in[3]$.

To ensure server-privacy, define an interpolation polynomial $\psi^s(\alpha)$ of degree $G(K+X-1)+E+T-1=17$ such that
\begin{IEEEeqnarray}{rCll}
\psi^s(\beta_{\ell,k})&=&0,&\quad\forall\, \ell\in[3],k\in[2s-1:2s],\label{example:user privacy2}\\
\psi^s(\alpha_{j})&=&z_{j}^{s},&\quad\forall\, j\in[12],  \label{example:user privacy}
\end{IEEEeqnarray}
where $z_{1}^{s},\ldots,z_{12}^{s}$ are the random variables shared over the servers.

Let $\zeta^s(\alpha)$ be the response polynomial of degree $\deg(\zeta^s(\alpha))=17$ in round $s$:
\begin{IEEEeqnarray}{rCl}\label{example:24}
\zeta^s(\alpha)&=&\rho_{1}^{s}(\varphi_1^{(1)}(\alpha),\varphi_{1}^{(2)}(\alpha),\alpha)+\rho_{2}^{s}(\varphi_2^{(1)}(\alpha),\varphi_{2}^{(2)}(\alpha),\alpha)+\rho_{3}^{s}(\varphi_3^{(1)}(\alpha),\varphi_{3}^{(2)}(\alpha),\alpha)+\psi^s(\alpha).
\end{IEEEeqnarray}
Then, any authentic server $n$ responds the user with $A^s_n=\sum_{i\in[3]}\rho_{i}^{s}(\varphi_i^{(1)}(\alpha_n),\varphi_{i}^{(2)}(\alpha_n),\alpha_n)+\psi^s(\alpha_n)$, which is equivalent to evaluating $\zeta^s(\alpha)$ at $\alpha=\alpha_n$.
Remarkably, $(\zeta^s(\alpha_1),\ldots,\zeta^s(\alpha_{21}))$ is a $(21,18)$ RS codeword, which is robust against any $B=1$ Byzantine error and $U=1$ unresponsive error. Hence, the user can decode $\zeta^s(\alpha)$ from the answers of responsive servers by using RS decoding algorithms. 
Then, by \eqref{example:storage}-\eqref{example:user privacy}, evaluating the polynomial at $\alpha=\beta_{1,2s-1},\beta_{2,2s-1},\beta_{3,2s-1},\beta_{1,2s},\beta_{2,2s},\beta_{3,2s}$ can obtain
\begin{IEEEeqnarray*}{rClrCl}
\zeta^s(\beta_{1,2s-1})&=&v_{1,2s-1}^{(\theta)},\quad \zeta^s(\beta_{1,2s})&=&v_{1,2s}^{(\theta)},\notag \\
\zeta^s(\beta_{2,2s-1})&=&v_{2,2s-1}^{(\theta)},\quad \zeta^s(\beta_{2,2s})&=&v_{2,2s}^{(\theta)},\notag \\
\zeta^s(\beta_{3,2s-1})&=&v_{3,2s-1}^{(\theta)},\quad \zeta^s(\beta_{3,2s})&=&v_{3,2s}^{(\theta)}. \notag
\end{IEEEeqnarray*}
Finally, the user can recover the desired function evaluations in \eqref{example2:desired} after two rounds.
The scheme achieves the PPC rate $R_p=\frac{3}{10}$ and secrecy rate $R_s=2$.


\section{Feasibility and Performance of the Proposed U-B-MDS-XTSPPC Scheme}\label{proof:privacy2}
In this section, we show the feasibility of U-B-MDS-XTSPPC scheme and analyse its performance. 
Before that, some lemmas are presented, which will be used for analysing the arithmetic complexities of the proposed scheme.
\begin{Lemma}[Polynomial Evaluation and Interpolation \cite{Von}]\label{Lamma:complexity11}
The evaluation of a $k$-th degree polynomial at $k+1$ arbitrary points can be done in ${\mathcal{O}}(k(\log k)^2\log\log k)$ arithmetic operations, and consequently, its dual problem, interpolation of a $k$-th degree polynomial from $k+1$ arbitrary points can be performed in the same arithmetic operations ${\mathcal{O}}(k(\log k)^2\log\log k)$.
\end{Lemma}

\begin{Lemma}[Multivariate Polynomial Evaluation \cite{Ballico,Lodha}]\label{Lamma:complexity22}
The evaluation of a multivariate polynomial of degree $n$ in $k$ variables can be done in $\mathcal{O}(n^k)$ arithmetic operations.
\end{Lemma}

\begin{Lemma}[Decoding Reed-Solomon Codes \cite{Lin,Gao}]\label{Lamma:complexity33}
Decoding Reed-Solomon codes of dimension $n$ with $b$ errors and $u$ erasures over arbitrary finite fields can be done in $\mathcal{O}(n(\log n)^2\log\log n)$ arithmetic operations by utilizing fast polynomial multiplications \cite{Von} if its minimum distance satisfies $d>2b+u$.
\end{Lemma}

\begin{Theorem}\label{lemma:achievability}
The proposed PPC scheme in Section \ref{s-PPC scheme} is robust against $X$-secure data storage, $B$ Byzantine and $U$ unresponsive servers, $T$-colluding user-privacy, and server-privacy.
\end{Theorem}
\begin{IEEEproof}
It is sufficient to prove that the U-B-MDS-XTSPPC scheme satisfies the constraints of \eqref{X security}, \eqref{correct:}, \eqref{Infor:priva cons}, and \eqref{Infor:priva cons2}.

\subsubsection*{$X$-Security}
For any subset $\{n_1,\ldots,n_X\}\subseteq[N]$ of size $X$, let
\begin{IEEEeqnarray}{c}\notag
\mathbf{C}_\ell=
\left[
\begin{array}{cccc}
  c_{\ell,1}(\alpha_{n_1}) & c_{\ell,2}(\alpha_{n_1}) & \ldots & c_{\ell,X}(\alpha_{n_1}) \\
  c_{\ell,1}(\alpha_{n_2}) & c_{\ell,2}(\alpha_{n_2}) & \ldots & c_{\ell,X}(\alpha_{n_2}) \\
  \vdots & \vdots & \ddots & \vdots \\
  c_{\ell,1}(\alpha_{n_X}) & c_{\ell,2}(\alpha_{n_X}) & \ldots & c_{\ell,X}(\alpha_{n_X}) \\
\end{array}
\right],\quad\forall\,\ell\in[L],
\end{IEEEeqnarray}
where
\begin{IEEEeqnarray}{rCl}\notag
c_{\ell,i}(\alpha)&=&\left(\prod\limits_{j\in[K]}\frac{\alpha-\beta_{\ell,j}}{\beta_{\ell,K+i}-\beta_{\ell,j}}\right)\cdot\prod\limits_{j\in[K+1:K+X]\backslash\{K+i\}}\frac{\alpha-\beta_{\ell,j}}{\beta_{\ell,K+i}-\beta_{\ell,j}},\quad\forall\,i\in[X].
\end{IEEEeqnarray}

Thus, by applying P1-P4 to Lemma \ref{g-cauchy matrix}, $\mathbf{C}_\ell$ is a generalized Cauchy matrix and is non-singular over $\mathbb{F}_q$ for any $\ell\in[L]$.
Then,  according to \eqref{storage code} and Lemma \ref{lemma:security},
\begin{IEEEeqnarray}{c}\notag
I(\varphi_\ell^{(m)}(\alpha_{n_1}),\ldots,\varphi_\ell^{(m)}(\alpha_{n_X});w_{\ell,1}^{(m)},\ldots,w_{\ell,K}^{(m)})=0, \quad\forall\, m\in[M],\ell\in[L].
\end{IEEEeqnarray}
Therefore,
\begin{IEEEeqnarray*}{rCl}
I(\mathbf{y}_{n_1},\ldots,\mathbf{y}_{n_X};\mathbf{W}^{(1)},\ldots,\mathbf{W}^{(M)})&\overset{(a)}{=}&I(\{\varphi_\ell^{(m)}(\alpha_{n_1}),\ldots,\varphi_\ell^{(m)}(\alpha_{n_X})\}_{\ell\in[L],m\in[M]};\{w_{\ell,1}^{(m)},\ldots,w_{\ell,K}^{(m)}\}_{\ell\in[L],m\in[M]} )\label{security:1}\\
&\overset{(b)}{=}&\sum\limits_{m\in[M]}\sum\limits_{\ell\in[L]}I(\varphi_\ell^{(m)}(\alpha_{n_1}),\ldots,\varphi_\ell^{(m)}(\alpha_{n_X});w_{\ell,1}^{(m)},\ldots,w_{\ell,K}^{(m)})\label{security:2}\\
&=&0,\label{security:3}
\end{IEEEeqnarray*}
where $(a)$ is due to \eqref{file symbols} and  \eqref{stored data}; $(b)$ follows from the fact that all the symbols $\{w_{\ell,1}^{(m)},\ldots,w_{\ell,K}^{(m)},z_{\ell,K+1}^{(m)},\ldots,z_{\ell,K+X}^{(m)}\}_{\ell\in[L],m\in[M]}$ are generated independently and uniformly from $\mathbb{F}_q$ and thus the sets of variables $\{\varphi_\ell^{(m)}(\alpha_{n_1}),\ldots,\varphi_\ell^{(m)}(\alpha_{n_X}),w_{\ell,1}^{(m)},\ldots,w_{\ell,K}^{(m)}\}$ are independent across all $m\in[M],\ell\in[L]$.  Therefore, $X$-security of the data files follows from \eqref{X security}.

\subsubsection*{Byzantine and Unresponsiveness}
By \eqref{decoding each round}, the user can recover all the desired evaluations correctly. Thus, the scheme can resist any $B$ Byzantine servers and $U$ unresponsive servers even if their identities change from round to round.
Accordingly, \eqref{correct:} follows.

\subsubsection*{User-Privacy}

Let $\mathcal{T}=\{n_1,n_2,\ldots,n_T\}\subseteq[N]$ be any $T$ indices of the $N$ servers.
For any $i\in[L]$, by \eqref{query:2}-\eqref{query:22}, the query elements $\rho_{i}^{s}(x_1,\ldots,x_M,\alpha_{n_1}),\rho_{i}^{s}(x_1,\ldots,x_M,\alpha_{n_2}),\ldots,\rho_{i}^{s}(x_1,\ldots,x_M,\alpha_{n_T})$ sent to the servers $\mathcal{T}$ are protected by $T$ random noise polynomials $\phi_{i,1}^{s},\ldots,\phi_{i,T}^{s}$ chosen independently and uniformly from the polynomial space $\mathcal{P}$, as shown below.
\begin{IEEEeqnarray}{c}\notag
\left[
\begin{array}{@{}c@{}}
  \rho_{i}^{s}(x_1,\ldots,x_M,\alpha_{n_1}) \\
  \rho_{i}^{s}(x_1,\ldots,x_M,\alpha_{n_2}) \\
  \vdots \\
  \rho_{i}^{s}(x_1,\ldots,x_M,\alpha_{n_T})
\end{array}
\right]=
\underbrace{\left[
\begin{array}{@{}c@{}}
  h(\alpha_{n_1}) \\
  h(\alpha_{n_2}) \\
  \vdots \\
  h(\alpha_{n_T})
\end{array}
\right]}_{=\mathbf{h}}+
\underbrace{
\left[
\begin{array}{cccc}
  f_1(\alpha_{n_1}) & f_2(\alpha_{n_1}) & \ldots & f_T(\alpha_{n_1}) \\
  f_1(\alpha_{n_2}) & f_2(\alpha_{n_2}) & \ldots & f_T(\alpha_{n_2}) \\
  \vdots & \vdots & \ddots & \vdots \\
  f_1(\alpha_{n_T}) & f_2(\alpha_{n_T}) & \ldots & f_T(\alpha_{n_T}) \\
\end{array}
\right]}_{\triangleq{\mathbf{F}}^s}
\underbrace{\left[
\begin{array}{@{}c@{}}
  \phi_{i,1}^{s} \\
  \phi_{i,2}^{s} \\
  \vdots \\
  \phi_{i,T}^{s}
\end{array}
\right]}_{=\bm{\phi}_i^s},
\end{IEEEeqnarray}
where
\begin{IEEEeqnarray}{c}\notag
h(\alpha)=\sum\limits_{l\in[(s-1)\Delta+1:s\Delta]}\phi^{(\theta)}\cdot\left(\prod\limits_{\substack{j\in[L],r\in[(s-1)\Delta+1:s\Delta]\\(j,r)\neq(i,l)}}\frac{\alpha-\beta_{j,r}}{\beta_{i,l}-\beta_{j,r}}\right)
\left(\prod\limits_{v\in[T]}\frac{\alpha-\alpha_v}{\beta_{i,l}-\alpha_v}\right),
\end{IEEEeqnarray}
and
\begin{IEEEeqnarray}{c}\notag
f_l(\alpha)=\left(\prod\limits_{j\in[L],r\in[(s-1)\Delta+1:s\Delta]}\frac{\alpha-\beta_{j,r}}{\alpha_{l}-\beta_{j,r}}\right)\left(\prod\limits_{v\in[T]\backslash\{l\}}\frac{\alpha-\alpha_v}{\alpha_{l}-\alpha_v}\right),\quad\forall\,l\in[T].
\end{IEEEeqnarray}

According to P3 and P4 again, the elements $\alpha_1,\ldots,\alpha_N$ are distinct and $\{\alpha_n:n\in[N]\}\cap\{\beta_{j,r}:j\in[L],r\in[(s-1)\Delta+1:s\Delta]\}=\emptyset$ for any $s\in[S]$. Hence, $\mathbf{F}^{s}$ is invertible by Lemma \ref{g-cauchy matrix}, whose inverse matrix is denoted by $(\mathbf{F}^{s})^{-1}$.
Then,
\begin{IEEEeqnarray}{rCl}
&&I\big(\{\rho_{i}^{s}(x_1,\ldots,x_M,\alpha_{n_1}),\ldots,\rho_{i}^{s}(x_1,\ldots,x_M,\alpha_{n_T})\}_{i\in[L],s\in[S]};\theta\big)\notag\\
&\overset{(a)}{=}&I\big(\{\rho_{i}^{s}(x_1,\ldots,x_M,\alpha_{n_1}),\ldots,\rho_{i}^{s}(x_1,\ldots,x_M,\alpha_{n_T})\}_{i\in[L],s\in[S]};\phi^{(\theta)}\big)\notag\\
&=&I\big(\{\mathbf{h}+\mathbf{F}^{s}\cdot\bm{\phi}_i^s\}_{i\in[L],s\in[S]};\phi^{(\theta)}\big) \label{privacy:111} \\
&=&I\big(\{\left(\mathbf{F}^{s}\right)^{-1}\cdot\mathbf{h}+\bm{\phi}_i^s\}_{i\in[L],s\in[S]};\phi^{(\theta)}\big) \\
&=&H\big(\{\left(\mathbf{F}^{s}\right)^{-1}\cdot\mathbf{h}+\bm{\phi}_i^s\}_{i\in[L],s\in[S]}\big)-
H\big(\{\left(\mathbf{F}^{s}\right)^{-1}\cdot\mathbf{h}+\bm{\phi}_i^s\}_{i\in[L],s\in[S]}|\phi^{(\theta)}\big) \\
&\overset{(b)}{=}&H\big(\{\left(\mathbf{F}^{s}\right)^{-1}\cdot\mathbf{h}+\bm{\phi}_i^s\}_{i\in[L],s\in[S]}\big)-
H\big(\{\left(\mathbf{F}^{s}\right)^{-1}\cdot\mathbf{h}+\bm{\phi}_i^s\}_{i\in[L],s\in[S]}\big)\\
&=&0,  \label{privacy:222}
\end{IEEEeqnarray}
where $(a)$ is due to the fact that the candidate polynomials $\{\phi^{(u)}\}_{u\in[P]}$ are globally known and thus $\phi^{(\theta)}$ and $\theta$ are determined of each other;
$(b)$ is due to the fact that $\bm{\phi}_i^s$ are independently and uniformly distributed on the polynomial space $\mathcal{P}$ spanned by the candidate functions $\{\phi^{(u)}\}_{u\in[P]}$ defined over $\mathbb{F}_q$ and are generated independently of $(\mathbf{F}^{s})^{-1}\cdot\mathbf{h},\phi^{(\theta)}$ for all $i\in[L],s\in[S]$, and thus
$H\big(\{\left(\mathbf{F}^{s}\right)^{-1}\cdot\mathbf{h}+\bm{\phi}_i^s\}_{i\in[L],s\in[S]}|\phi^{(\theta)}\big)=
H\big(\{\left(\mathbf{F}^{s}\right)^{-1}\cdot\mathbf{h}+\bm{\phi}_i^s\}_{i\in[L],s\in[S]}\big)$.

Further, by \eqref{query:22},
\begin{IEEEeqnarray*}{rCl}
I\big(\{\mathbf{q}_{n_1}^{s},\ldots,\mathbf{q}_{n_T}^{s}\}_{s\in[S]};\theta\big)=I\big(\{\rho_{i}^{s}(x_1,\ldots,x_M,\alpha_{n_1}),\ldots,\rho_{i}^{s}(x_1,\ldots,x_M,\alpha_{n_T})\}_{i\in[L],s\in[S]};\theta\big)
=0.
\end{IEEEeqnarray*}
Thus, \eqref{Infor:priva cons} is proved.

\subsubsection*{Server-Privacy}
For any $\mathcal{B}^{s}\subseteq[N],\mathcal{U}^{s}\subseteq[N],|\mathcal{B}^{s}|\leq B,|\mathcal{U}^{s}|\leq U,\mathcal{B}^{s}\cap\mathcal{U}^{s}=\emptyset$ such that $s\in[S]$, the identifies of which are unknown to the user, let $\mu^s(\alpha)$ be the first term of the answer polynomial $\zeta^s(\alpha)$ in \eqref{polynomial:scheme2}, i.e.,
\begin{IEEEeqnarray}{c}\label{inner poly}
\mu^s(\alpha)=\sum\limits_{i=1}^{L}\rho_{i}^{s}(\varphi_i^{(1)}(\alpha),\ldots,\varphi_{i}^{(M)}(\alpha),\alpha).
\end{IEEEeqnarray}

Then,  we have
\begin{IEEEeqnarray}{rCl}
0&\leq&I(\{A_{[N]\backslash\mathcal{U}^{s}}^{s},\mathbf{q}_{[N]}^{s}\}_{s\in[S]};\mathbf{W}^{(1)},\ldots,\mathbf{W}^{(M)}|\mathbf{V}^{(\theta)})\notag\\
&=&I(\{\mathbf{q}_{[N]}^{s}\}_{s\in[S]};\mathbf{W}^{(1)},\ldots,\mathbf{W}^{(M)}|\mathbf{V}^{(\theta)})+
I(\{A_{[N]\backslash\mathcal{U}^{s}}^{s}\}_{s\in[S]};\mathbf{W}^{(1)},\ldots,\mathbf{W}^{(M)}|\mathbf{V}^{(\theta)},\{\mathbf{q}_{[N]}^{s}\}_{s\in[S]})\notag\\
&\overset{(a)}{=}&I(\{A_{[N]\backslash\mathcal{U}^{s}}^{s}\}_{s\in[S]};\mathbf{W}^{(1)},\ldots,\mathbf{W}^{(M)}|\mathbf{V}^{(\theta)},\{\mathbf{q}_{[N]}^{s}\}_{s\in[S]})\notag\\
&\overset{(b)}{=}&I(\{A_{[N]\backslash\mathcal{U}^{s}}^{s},\zeta^{s}(\alpha_1),\ldots,\zeta^{s}(\alpha_N)\}_{s\in[S]};\mathbf{W}^{(1)},\ldots,\mathbf{W}^{(M)}|\mathbf{V}^{(\theta)},\{\mathbf{q}_{[N]}^{s}\}_{s\in[S]})\notag\\
&=&I(\{\zeta^{s}(\alpha_1),\ldots,\zeta^{s}(\alpha_N)\}_{s\in[S]};\mathbf{W}^{(1)},\ldots,\mathbf{W}^{(M)}|\mathbf{V}^{(\theta)},\{\mathbf{q}_{[N]}^{s}\}_{s\in[S]})\notag\\
&&+I(\{A_{[N]\backslash\mathcal{U}^{s}}^{s}\}_{s\in[S]};\mathbf{W}^{(1)},\ldots,\mathbf{W}^{(M)}|\mathbf{V}^{(\theta)},\{\mathbf{q}_{[N]}^{s}\}_{s\in[S]},\{\zeta^{s}(\alpha_1),\ldots,\zeta^{s}(\alpha_N)\}_{s\in[S]})\notag\\
&\overset{(c)}{=}&I(\{\zeta^{s}(\alpha_1),\ldots,\zeta^{s}(\alpha_N)\}_{s\in[S]};\mathbf{W}^{(1)},\ldots,\mathbf{W}^{(M)}|\mathbf{V}^{(\theta)},\{\mathbf{q}_{[N]}^{s}\}_{s\in[S]})\notag\\
&&+I(\{A_{\mathcal{B}^{s}}^{s}\}_{s\in[S]};\mathbf{W}^{(1)},\ldots,\mathbf{W}^{(M)}|\mathbf{V}^{(\theta)},\{\mathbf{q}_{[N]}^{s}\}_{s\in[S]},\{\zeta^{s}(\alpha_1),\ldots,\zeta^{s}(\alpha_N)\}_{s\in[S]})\notag\\
&\overset{(d)}{=}&I(\{\zeta^{s}(\alpha_1),\ldots,\zeta^{s}(\alpha_N)\}_{s\in[S]};\mathbf{W}^{(1)},\ldots,\mathbf{W}^{(M)}|\mathbf{V}^{(\theta)},\{\mathbf{q}_{[N]}^{s}\}_{s\in[S]})\notag\\
&\overset{(e)}{=}&I(\{\zeta^{s}(\alpha):\alpha\in\{\alpha_j\}_{j\in[G(K+X-1)+T]}\cup\{\beta_{\ell,k}\}_{\ell\in[L],k\in[(s-1)\Delta+1:s\Delta]}\}_{s\in[S]};\mathbf{W}^{(1)},\ldots,\mathbf{W}^{(M)}|\mathbf{V}^{(\theta)},\{\mathbf{q}_{[N]}^{s}\}_{s\in[S]}) \notag\\
&\overset{(f)}{=}&I(\{v_{\ell,k}^{(\theta)}\}_{\ell\in[L],k\in[(s-1)\Delta+1:s\Delta],s\in[S]},\{\mu^s(\alpha_j)+z_j^{s}\}_{j\in[G(K+X-1)+T],s\in[S]};\mathbf{W}^{(1)},\ldots,\mathbf{W}^{(M)}|\mathbf{V}^{(\theta)},\{\mathbf{q}_{[N]}^{s}\}_{s\in[S]}) \notag\\
&\overset{(g)}{=}&I(\{\mu^s(\alpha_j)+z_j^{s}\}_{j\in[G(K+X-1)+T],s\in[S]};\mathbf{W}^{(1)},\ldots,\mathbf{W}^{(M)}|\mathbf{V}^{(\theta)},\{\mathbf{q}_{[N]}^{s}\}_{s\in[S]})\label{terms}\\
&\overset{(h)}{=}&0,\notag
\end{IEEEeqnarray}
where $(a)$ follows from the fact that the queries $\{\mathbf{q}_{[N]}^{s}\}_{s\in[S]}$ are determined by $\phi^{(\theta)}$ and the random noise polynomials $\{\phi_{i,t}^{s}\}_{i\in[L],t\in[T]}$ that  are generated independently of the data $\mathbf{W}^{(1)},\ldots,\mathbf{W}^{(M)},\mathbf{V}^{(\theta)}$ by \eqref{query:2}-\eqref{query:22}, thus
\begin{IEEEeqnarray*}{rCl}
0=I(\phi^{(\theta)},\{\phi_{i,t}^{s}\}_{i\in[L],t\in[T]};\mathbf{W}^{(1)},\ldots,\mathbf{W}^{(M)},\mathbf{V}^{(\theta)})
\geq I(\{\mathbf{q}_{[N]}^{s}\}_{s\in[S]};\mathbf{W}^{(1)},\ldots,\mathbf{W}^{(M)}|\mathbf{V}^{(\theta)})\geq0;
\end{IEEEeqnarray*}
$(b)$ holds because the user can decode the polynomial $\zeta^s(\alpha)$ in \eqref{polynomial:scheme2} from the answers $A_{[N]\backslash\mathcal{U}^{s}}^{s}$ even if there exists $B$ Byzantine errors for any $s\in[S]$;
$(c)$ is due to the fact that the answer $A_{n}^{s}$ is equivalent to $\zeta^{s}(\alpha_n)$ for any authentic server $n\in[N]\backslash(\mathcal{U}^{s}\cup\mathcal{B}^{s})$ by \eqref{answer:22} and \eqref{polynomial:scheme2};
$(d)$ follows by the fact \cite{Tajeddine222,Wang123,wang-byzantine} that the Byzantine servers return arbitrary responses $\{A_{\mathcal{B}^{s}}^{s}\}_{s\in[S]}$ maliciously to confuse the user 
and thus the answers of Byzantine servers cannot leak anything about the files $\mathbf{W}^{(1)},\ldots,\mathbf{W}^{(M)}$ to the user;
$(e)$ holds because $\zeta^{s}(\alpha)$ is a polynomial of degree $G(K+X-1)+T+L\Delta-1$ such that $\{\zeta^{s}(\alpha_1),\ldots,\zeta^{s}(\alpha_N)\}$ and $\{\zeta^{s}(\alpha):\alpha\in\{\alpha_j\}_{j\in[G(K+X-1)+T]}\cup\{\beta_{\ell,k}\}_{\ell\in[L],k\in[(s-1)\Delta+1:s\Delta]}\}$ are determined of each other by Lagrange interpolation rules and P2-P4 for any $s\in[S]$;
$(f)$ follows by \eqref{symmetric:122}, \eqref{answer:22}, \eqref{evaluating:1} and \eqref{inner poly};
$(g)$ is due to \eqref{desired file};
$(h)$ follows from the fact that $\{z_j^{s}\}_{j\in[G(K+X-1)+T],s\in[S]}$ are i.i.d. uniformly over $\mathbb{F}_q$ and are generated independently of all other variables in \eqref{terms}.
This completes the proof of \eqref{Infor:priva cons2}.
\end{IEEEproof}

The performance of U-B-MDS-XTSPPC scheme is characterized in the following theorem.

\begin{Theorem}\label{results:schemes1}
Given any candidate polynomial functions $\{\phi^{(u)}\}_{u\in[P]}$, let $ F $ be the dimension of the vector space spanned by the candidate polynomial functions $\{\phi^{(u)}\}_{u\in[P]}$ over $\mathbb{F}_q$. If $N>G(K+X-1)+T+2B+U$, the U-B-MDS-XTSPPC scheme using Lagrange encoding achieves
\begin{IEEEeqnarray*}{rCl}
\text{PPC Rate:}\quad& R_p=\frac{E}{N-U}, \\
\text{Secrecy Rate:}\quad& R_s=\frac{G(K+X-1)+T}{E}, \\
\text{Upload Cost:}\quad& C_u=\frac{KNE F }{(\gcd(K,E))^2},\\
\text{Finite Field Size:}\quad&  q\geq N+\max\{K,E\}, \\
\text{Query Complexity:}\quad& \mathcal{C}_q=\mathcal{O}\left(\frac{KE F  N(\log N)^2\log\log N}{(\gcd(K,E))^2}\right), \\
\text{Server Computation Complexity:}\quad& \mathcal{C}_s=\mathcal{O}\left(\frac{KEG^M}{(\gcd(K,E))^2}\right), \\
\text{Decoding Complexity:}\quad& \mathcal{C}_d=\mathcal{O}\left(\frac{KN(\log N)^2\log\log N}{\gcd(K,E)}\right),
\end{IEEEeqnarray*}
where $E=N-(G(K+X-1)+T+2B+U)$.
\end{Theorem}

\begin{IEEEproof}
The performance of the proposed scheme is analyzed as follows.
\subsubsection*{PPC Rate and Secrecy Rate}
In each round, the user downloads $N-U$ symbols from the responsive servers by \eqref{answer:22}. Thus, the PPC rate \eqref{def:PPC rate} is
\begin{IEEEeqnarray}{c}\label{rate computation}
R_p=\frac{LK}{D}=\frac{L K}{\sum_{s=1}^{S}(N-U)}=\frac{E}{N-U}.
\end{IEEEeqnarray}

By \eqref{common:random}, each round requires the number of common random variables to be $G(K+X-1)+T$. Thus, the security rate \eqref{def:Secrecy rate} is
\begin{IEEEeqnarray}{c}\notag
R_s=\frac{H(\mathcal{F})}{LK}=\frac{\sum_{s=1}^{S}(G(K+X-1)+T)}{LK}=\frac{G(K+X-1)+T}{E}.
\end{IEEEeqnarray}

\subsubsection*{Upload Cost}
By \eqref{query:22}, the query sent to each server in each round is composed of  $L$ polynomials belonging to the polynomial space $\mathcal{P}$ spanned by the candidate functions $\{\phi^{(u)}\}_{u\in[P]}$ over $\mathbb{F}_q$.
Since the function set $\mathcal{P}$ is a polynomial space defined over $\mathbb{F}_q$, there exists a \emph{bijection} between the function set and a vector space of dimension $ F =\log_q|\mathcal{P}|$. Consequently, each polynomial function in the set can be denoted by a vector of length $ F $.
Hence, the upload cost is $C_u=SNL F $.

\subsubsection*{Finite Field Size} By Lemma \ref{theorem:tield size},  the finite field $\mathbb{F}_q$ is enough with size $q\geq N+\max\{K, E\}$.

\subsubsection*{System Complexity}
For query complexity,  based on the bijection above, the queries \eqref{query:22} sent to servers can be viewed as evaluating $L$ polynomials of degree less than $N$ at $N$ points for $ F $ times in each round. So, the queries achieve the complexity at most $\mathcal{O}(SL F  N(\log N)^2\log\log N)$ by Lemma \ref{Lamma:complexity11}.

For the computation complexity at servers \eqref{answer:22}, each server first evaluates $L$ polynomials of degree at most $G$ in $M$ variables at distinct points and evaluates $\psi^{s}(\alpha)$ at one point, and then responds the sum of these polynomial evaluations for $S$ rounds. By Lemma \ref{Lamma:complexity22}, the complexity of evaluating the multivariate polynomials is at most $\mathcal{O}(SL G^M)$, which dominates the server computation complexity.

For decoding complexity, in each round, the user first decodes the answer polynomial $\zeta^s(\alpha)$ from a RS codeword of dimension $N$ and then evaluates the polynomial at $L\Delta=E$ ($E<N$) points. By Lemmas \ref{Lamma:complexity33} and \ref{Lamma:complexity11}, such operations of RS decoding and evaluations can be done in the complexity $\mathcal{O}(N(\log N)^2\log\log N)$.
Thus, the decoding complexity is $\mathcal{O}(SN(\log N)^2\log\log N)$ for $S$ rounds.
\end{IEEEproof}

\begin{Remark}
To the best of our knowledge, the download cost ($D=LK/R_p$) from responsive servers, the amount of random variables shared among the servers ($H(\mathcal{F})=LK R_s$), server computation complexity and decoding complexity grow with the size of data file in all the known PPC/PLC/PIR schemes (for examples, \cite{Raviv PPC,PC1,Sun replicated,Ulukus_MDS,Zhu,S-PIR1,Wang-MDS-SPIR,wang-colluding-SPIR}), however this is not the case for upload cost, finite field size and query complexity.
For example, assume the size of original file is enlarged by a factor of $\kappa$ for any positive integer $\kappa$. By convenience, we think of each `symbol' in the original data files as a `chunk' of length $\kappa$ symbols. To retrieve the desired data information, the user can still use the current PPC/PLC/PIR schemes such that the queries, answers and decoding happen over the chunks. Hence, the download cost, the amount of random variables shared, server computation complexity and decoding complexity are scaled with $\kappa$, but the upload cost, finite field size and query complexity are independent of $\kappa$ since the original queries can be reused $\kappa$ times, each for  all the chunks.
\end{Remark}

\begin{Remark}
When server-privacy is not considered, U-B-MDS-XTSPPC automatically degrades to the setup in \cite{Raviv PPC}.
As we mentioned in Remark 2,  the two schemes in \cite{Raviv PPC} and this paper employed distinct coding techniques to align the interference.
Actually, the PPC scheme in \cite{Raviv PPC} first resorts to the interference alignment ideas of U-B-MDS-TPIR scheme in \cite{Tajeddine222} to retrieve all the coefficients of some composite polynomial functions, then recovers the composite polynomials, and finally evaluates the composite polynomials to obtain the desired evaluations.
Whereas in our PPC scheme,  we use Lagrange interpolation polynomials with multiple groups of distinct coding parameters to create interference alignment opportunities, such that the user can distinguish all the desired function evaluations and thus retrieves the desired evaluations directly from the server responses, see \eqref{evaluating:12345}--\eqref{evaluating:1}.
Consequently, our PPC scheme and the scheme in \cite{Raviv PPC}  can retrieve the same number of desired evaluations and the coefficients of the composite polynomial functions from the server responses of each round, respectively.
But the number of the coefficients of the composite polynomial functions is much large than the number of desired evaluations, thus our scheme achieves lower communication cost.
Accordingly, the PPC rate $\frac{N-(G(K+X-1)+T+2B+U)}{N-U}$ \eqref{rate computation} of our scheme avoids the penalty factor of $\frac{K}{G(K+X-1)+1}$, compared to the rate $\frac{N-(G(K+X-1)+T+2B+U)}{N-U}\cdot\frac{K}{G(K+X-1)+1}$ (in Equation (1) in \cite{Raviv PPC}) of the scheme \cite{Raviv PPC}.

\end{Remark}

\section{Performance Comparison}\label{sec:comparison}
In this section, we demonstrate the performance comparisons between the proposed U-B-MDS-XTSPPC scheme and the existing MDS-PPC schemes.
To the best of our knowledge, the known MDS-PPC schemes are  the MDS-PPC scheme \cite{PC5}, systematic MDS-PPC scheme \cite{PC5}, systematic MDS-TPPC scheme \cite{PC4}, and U-B-MDS-XTPPC scheme \cite{Raviv PPC}.
Particularly, the U-B-MDS-XTSPPC setup includes the setups of these schemes \cite{PC5,PC4,Raviv PPC} as special cases.
Notice that, the upload cost, finite field size and system complexity were not considered to measure the performance of PPC schemes in the previous setups, i.e., the metric they focus on is just PPC rate.
Therefore, for comparison and completeness, we conduct the performance analyses of PPC schemes in \cite{PC4,PC5,Raviv PPC} and list them in Tables \ref{U-B-MDS-XTPPC-1}, \ref{U-B-MDS-XTPPC-2} and \ref{U-B-MDS-XTPPC-3} for the setups of U-B-MDS-XTPPC, MDS-TPPC and MDS-PPC, respectively, where $ F $ is the dimension of the vector space spanned by the candidate polynomial functions $\{\phi^{(u)}\}_{u\in[P]}$ over $\mathbb{F}_q$.
%

\begin{table*}[htbp]
\renewcommand{\arraystretch}{1.25}
\centering
\caption{Performance comparison for U-B-MDS-XTPPC problem} \label{U-B-MDS-XTPPC-1}
\begin{threeparttable}
  \begin{tabular}{|c|c|c|}
  \hline
   & PPC Rate & Upload Cost \\ \hline
  U-B-MDS-XTPPC Scheme \cite{Raviv PPC} & $\frac{E}{N-U}\cdot\frac{K}{G(K+X-1)+1}$ & $(G(K+X-1)+1)N F $  \\ \hline
 Our Degraded U-B-MDS-XTPPC Scheme & $\frac{E}{N-U}$ & $\frac{KNE F }{(\gcd(K,E))^2}$ \\ \hline\hline 
   &  Finite Field Size & Query Complexity \\ \hline
  U-B-MDS-XTPPC Scheme \cite{Raviv PPC} & $q\geq N+K$ & $\mathcal{O}\big(G(K+X) F  N(\log N)^2\log\log N\big)$ \\ \hline
Our  Degraded U-B-MDS-XTPPC Scheme & $q\geq N+\max\{K,E\}$ & $\mathcal{O}\left(\frac{KE F  N(\log N)^2\log\log N}{(\gcd(K,E))^2}\right)$  \\ \hline\hline 
   & Server Computation Complexity & Decoding Complexity \\ \hline
  U-B-MDS-XTPPC Scheme \cite{Raviv PPC}  &  $\mathcal{O}(G(K+X)G^{M})$ & $\mathcal{O}(G(K+X)N(\log N)^2\log\log N)$ \\ \hline
 Our Degraded U-B-MDS-XTPPC Scheme  & $\mathcal{O}\left(\frac{KEG^M}{(\gcd(K,E))^2}\right)$ & $\mathcal{O}\left(\frac{KN(\log N)^2\log\log N}{\gcd(K,E)}\right)$ \\ \hline 
  \end{tabular}
  \begin{tablenotes}
       \footnotesize
       \item[]Here, $E=N-(G(K+X-1)+T+2B+U)$.
     \end{tablenotes}
\end{threeparttable}
\end{table*}

\begin{table*}[htbp]
\centering
\caption{Performance comparison for MDS-TPPC problem} \label{U-B-MDS-XTPPC-2}
\renewcommand{\arraystretch}{1.25}
\begin{threeparttable}
  \begin{tabular}{|c|c|c|}
  \hline
  & PPC Rate & Upload Cost \\ \hline
  Systematic MDS-TPPC Scheme \cite{PC4} & $\frac{\min\{E,K\}}{N}$ & $\frac{KN F }{\gcd(K,\min\{K,E\})}$ \\ \hline
  Degraded MDS-TPPC Scheme \cite{Raviv PPC} & $\frac{E}{N}\cdot\frac{K}{G(K-1) +1}$ & $(G(K-1) +1)N F $  \\ \hline
  Our Degraded MDS-TPPC Scheme & $\frac{E}{N}$ & $\frac{KNE F }{(\gcd(K,E))^2}$ \\ \hline\hline 
  & Finite Field Size  & Query Complexity  \\ \hline
  Systematic MDS-TPPC Scheme \cite{PC4} & $q\geq N$ & $\mathcal{O}\left(\frac{NK F ^2}{\gcd(K,\min\{K,E\})}\right)$ \\ \hline
  Degraded MDS-TPPC Scheme \cite{Raviv PPC} & $q\geq N+K$ & $\mathcal{O}\left(GK  F  N(\log N)^2\log\log N\right)$ \\ \hline
  Our Degraded MDS-TPPC Scheme & $q\geq N+\max\{K,E\}$ & $\mathcal{O}\left(\frac{KE F  N(\log N)^2\log\log N}{(\gcd(K,E))^2}\right)$  \\ \hline\hline 
  & Server Computation Complexity & Decoding Complexity \\ \hline
  Systematic MDS-TPPC Scheme \cite{PC4} & $\mathcal{O}\left(\frac{KG^M}{\gcd(K,\min\{K,E\})}\right)$ & $\mathcal{O}\left(\frac{KN(\log N)^2\log\log N}{\gcd(K,\min\{K,E\})}\right)$ \\ \hline
  Degraded MDS-TPPC Scheme \cite{Raviv PPC}  &  $\mathcal{O}(GK G^{M})$ & $\mathcal{O}(GK N(\log N)^2\log\log N)$ \\ \hline
  Our Degraded MDS-TPPC Scheme  & $\mathcal{O}\left(\frac{KEG^M}{(\gcd(K,E))^2}\right)$ & $\mathcal{O}\left(\frac{KN(\log N)^2\log\log N}{\gcd(K,E)}\right)$ \\ \hline 
  \end{tabular}
  \begin{tablenotes}
       \footnotesize
       \item[]Note that $E=N-(G(K-1)+T)$.
     \end{tablenotes}
\end{threeparttable}
\end{table*}

Specifically, if server-privacy is not considered (see Remark \ref{remark:degrade}), our scheme can straightly degrade to the case of U-B-MDS-XTPPC. Table \ref{U-B-MDS-XTPPC-1} compares the degraded scheme with the U-B-MDS-XTPPC scheme \cite{Raviv PPC}.
If one further assumes $U=B=X=0$, the proposed U-B-MDS-XTSPPC scheme and U-B-MDS-XTPPC scheme \cite{Raviv PPC} degrade to the setup of MDS-TPPC, which are compared in Table \ref{U-B-MDS-XTPPC-2} together with the systematic MDS-TPPC scheme \cite{PC4}. Moreover, if we further set $T=1$, all these schemes degrade to the setup of MDS-PPC, which are compared with nonsystematic MDS-PPC scheme \cite{PC5} and systematic MDS-PPC scheme \cite{PC5} in Table \ref{U-B-MDS-XTPPC-3}. Notably, as the number of files $M\rightarrow\infty$, the achievable rate of the nonsystematic MDS-PPC scheme approaches the rate of \cite{Raviv PPC} asymptotically, whereas the systematic scheme achieves the MDS-PPC rate
\begin{IEEEeqnarray}{l}
      R_{\infty}'=
      \left\{\begin{array}{@{}ll}
                                \frac{N-{K'}}{N},&\mathrm{if}~\big\lfloor\frac{N}{{K'}}\big\rfloor=1 ~\mathrm{and}~N-\big\lfloor\frac{N}{{K'}}\big\rfloor{K'}<K \\
                                \frac{1}{K+\big(\lfloor\frac{N}{{K'}}\rfloor-1\big){K'}}\big( \big\lfloor\frac{N}{{K'}} \big\rfloor K-K\big),  &\mathrm{if}~\big\lfloor\frac{N}{{K'}}\big\rfloor >1 ~\mathrm{and}~N-\big\lfloor\frac{N}{{K'}}\big\rfloor{K'}<K \\
                                \frac{1}{K+\lfloor\frac{N}{{K'}}\rfloor{K'}}\big\lfloor\frac{N}{{K'}} \big\rfloor K,  &\mathrm{if}~\big\lfloor\frac{N}{{K'}}\big\rfloor\geq 1 ~\mathrm{and}~N-\big\lfloor\frac{N}{{K'}}\big\rfloor{K'}\geq K \\
                                \end{array}
      \right.,\notag
\end{IEEEeqnarray}
where ${K'}\triangleq G(K-1)+1$. 


For the three classes of PPC problems, the main differences between the proposed scheme and the existing schemes are outlined as follows.
\begin{enumerate}
\item As the main performance measure of the PPC problem, the PPC rates achieved by our degraded PPC schemes are strictly superior to the previous best known (asymptotic) rates for the three classes of PPC problems.
\item In terms of finite field size, our degraded PPC schemes \emph{slightly} increase the field size.
\item In terms of upload cost, query complexity, server computation complexity and decoding complexity, our degraded PPC schemes outperform all the listed schemes apart from Systematic MDS-TPPC Scheme \cite{PC4} and its Degraded Systematic MDS-PPC Scheme for the three PPC problems.
\item Particularly, the U-B-MDS-XTPPC scheme \cite{Raviv PPC} along with its Degraded MDS-TPPC Scheme and Degraded MDS-PPC Scheme require the queries, answers and decoding to happen over multi-rounds, and \emph{successive decoding} with interference cancellation strategy is employed by the user to recover the desired information.  
  However, our general scheme and its degraded schemes can be carried out \emph{independently and concurrently}, which greatly improves the efficiency of retrieving desired function evaluations.
\end{enumerate}

\begin{table*}[htbp]
\centering
\caption{Performance comparison for MDS-PPC problem} \label{U-B-MDS-XTPPC-3}
\renewcommand{\arraystretch}{1.25}
\begin{threeparttable}
  \begin{tabular}{|c|c|c|}
  \hline
   & (Asymptotic) PPC Rate &  Upload Cost  \\ \hline
  Nonsystematic MDS-PPC Scheme \cite{PC5} & $\frac{E}{N}\cdot\frac{K}{G(K-1)+1}$ & $\mathcal{O}\left(NP\log_q\frac{N^{P}!}{(N^{P-1}K')!}\right)$ \\ \hline
  Systematic MDS-PPC Scheme \cite{PC5} & $R_{\infty}'$ & $\mathcal{O}\left(N'P\log_q\frac{{E'}^{P}!}{({E'}^{P-1}K')!}\right)$ \\ \hline
  Degraded Systematic MDS-PPC Scheme \cite{PC4} & $\frac{\min\{E,K\}}{N}$ & $\frac{KN F }{\gcd(K,\min\{K,E\})}$ \\ \hline
  Degraded MDS-PPC Scheme \cite{Raviv PPC} & $\frac{E}{N}\cdot\frac{K}{G(K-1)+1}$ & $(G(K-1) +1)N F $  \\ \hline
 Our Degraded MDS-PPC Scheme & $\frac{E}{N}$ & $\frac{KNE F }{(\gcd(K,E))^2}$ \\ \hline\hline 
   & Finite Field Size & Query Complexity  \\ \hline
  Nonsystematic MDS-PPC Scheme \cite{PC5} & $q\geq N$ & $\mathcal{O}(NP K'N^{P-1})$ \\ \hline
  Systematic MDS-PPC Scheme \cite{PC5} & $q\geq N$ & $\mathcal{O}(NP K{E'}^{P-1})$ \\ \hline
  Degraded Systematic MDS-PPC Scheme \cite{PC4} & $q\geq N$ & $\mathcal{O}\left(\frac{NK F ^2}{\gcd(K,\min\{K,E\})}\right)$ \\ \hline
  Degraded MDS-PPC Scheme \cite{Raviv PPC} & $q\geq N+K$ & $\mathcal{O}\left(GK F  N(\log N)^2\log\log N\right)$ \\ \hline
  Our Degraded MDS-PPC Scheme & $q\geq N+\max\{K,E\}$ & $\mathcal{O}\left(\frac{KE F  N(\log N)^2\log\log N}{(\gcd(K,E))^2}\right)$  \\ \hline\hline 
   & Server Computation Complexity & Decoding Complexity \\ \hline
  Nonsystematic MDS-PPC Scheme \cite{PC5} &  $\mathcal{O}(N^{P}P G^{M})$ & $\mathcal{O}(KN^{P}N(\log N)^2\log\log N)$ \\ \hline
  Systematic MDS-PPC Scheme \cite{PC5} &  $\mathcal{O}({E'}^{P}P G^{M})$ & $\mathcal{O}(K{E'}^{P}{N'}(\log{N'})^2\log\log{N'})$   \\ \hline
  Degraded Systematic MDS-PPC Scheme \cite{PC4} & $\mathcal{O}\left(\frac{KG^M}{\gcd(K,\min\{K,E\})}\right)$ & $\mathcal{O}\left(\frac{KN(\log N)^2\log\log N}{\gcd(K,\min\{K,E\})}\right)$ \\ \hline
  Degraded MDS-PPC Scheme \cite{Raviv PPC}  &  $\mathcal{O}(GK G^{M})$ & $\mathcal{O}(GK N(\log N)^2\log\log N)$ \\ \hline
 Our Degraded MDS-PPC Scheme  & $\mathcal{O}\left(\frac{KEG^M}{(\gcd(K,E))^2}\right)$ & $\mathcal{O}\left(\frac{KN(\log N)^2\log\log N}{\gcd(K,E)}\right)$ \\ \hline 
  \end{tabular}
\begin{tablenotes}
       \footnotesize
       \item[]Here, $E=N-(G(K-1)+1)$, $E'={N'}-\big\lfloor\frac{{N'}}{{K'}}\big\rfloor({K'}-K)$,  \\ and ${N'}=
      \left\{\begin{array}{@{}ll}
                                N,&\mathrm{if}~\big\lfloor\frac{N}{{K'}}\big\rfloor=1 ~\mathrm{and}~N-\big\lfloor\frac{N}{{K'}}\big\rfloor{K'}<K \\
                                K+\left(\lfloor\frac{N}{{K'}}\rfloor-1\right){K'},  &\mathrm{if}~\big\lfloor\frac{N}{{K'}}\big\rfloor >1 ~\mathrm{and}~N-\big\lfloor\frac{N}{{K'}}\big\rfloor{K'}<K \\
                                K+\lfloor\frac{N}{{K'}}\rfloor{K'},  &\mathrm{if}~\big\lfloor\frac{N}{{K'}}\big\rfloor\geq 1 ~\mathrm{and}~N-\big\lfloor\frac{N}{{K'}}\big\rfloor{K'}\geq K \\
                                \end{array}
      \right..$
     \end{tablenotes}
\end{threeparttable}
\end{table*}

\section{Conclusion}\label{conclusion}
In this paper, we focused on designing U-B-MDS-XTSPPC schemes with PPC rate as high as possible, while keeping secrecy rate, upload cost, finite field size, query complexity, server computation complexity and decoding complexity as small/low as possible.
A general PPC scheme was proposed for any candidate polynomial set. The PPC scheme achieves the PPC rate $1-\frac{G(K+X-1)+T+2B}{N-U}$ with secrecy rate $\frac{G(K+X-1)+T}{N-(G(K+X-1)+T+2B+U)}$ and finite field size $N+\max\{K,N-(G(K+X-1)+T+2B+U)\}$. Notably, the proposed scheme can operate on all the previous MDS-PPC setups, and improved their (asymptotical) PPC rates.

U-B-MDS-XTSPPC generalizes the current PIR/PLC/PPC settings and the proposed U-B-MDS-XTSPPC scheme (asymptotically) achieves the current optimal schemes for various special cases including \cite{Sun replicated,Ulukus_MDS,MDS-TPIR,Tajeddine222,MDS-X-security,PC1,MDS Obead,PC3,S-PIR1,Wang-MDS-SPIR,wang-colluding-SPIR}.
Thus, we conjecture that the proposed U-B-MDS-XTSPPC scheme is also optimal.
Naturally, this raises two promising open problems. One is to prove the question of optimality of the proposed solution, and the other is to characterize the minimal amount of common randomness shared among servers for ensuring server privacy, which are valuable research directions for future work.

\begin{appendix}[Proof of Lemma \ref{theorem:tield size}]\label{public elements}
It is sufficient to design $\{\beta_{\ell,k},\alpha_n:\ell\in[L],k\in[K+X],n\in[N]\}\subseteq\mathbb{F}_q$ satisfying P1-P4 with
\begin{IEEEeqnarray}{c}\label{finite field size}
|\{\beta_{\ell,k},\alpha_n:\ell\in[L],k\in[K+X],n\in[N]\}|\leq \max\{K, E\}+N.
\end{IEEEeqnarray}

We generate these public elements in three steps.
\begin{basedescript}{\desclabelstyle{\pushlabel}\desclabelwidth{3em}}
  \item [Step 1.]  This step is done in two cases depending on $K\ge E$ or not.
  \begin{itemize}
  \item If $K\geq E$, let $(\beta_{1,1},\beta_{1,2},\ldots,\beta_{1,K})$ be $K$ distinct elements from $\mathbb{F}_q$. Then, set
  \begin{IEEEeqnarray}{c}\label{field:1}
  (\beta_{\ell,1},\beta_{\ell,2},\ldots,\beta_{\ell,K})=(\beta_{1,(\ell-1)\Delta+1},\ldots,\beta_{1,K},\beta_{1,1},\ldots,\beta_{1,(\ell-1)\Delta}),\quad\forall\,\ell\in[2:L ].
  \end{IEEEeqnarray}
  That is, the entries $(\beta_{\ell,1},\beta_{\ell,2},\ldots,\beta_{\ell,K})$ in row $\ell$ of the matrix $\boldsymbol{\beta}$ are given by performing left circular shift of the vector $(\beta_{1,1},\beta_{1,2},\ldots,\beta_{1,K})$ by $(\ell-1)\Delta$ positions for any $\ell\in[2:L ]$.
  \item If $K<E$, let $\{\beta_{\ell,k}:\ell\in[L],k\in[\Delta]\}$ be $L\Delta=E$ distinct elements from $\mathbb{F}_q$, i.e.,
  all the entries in the following sub-matrix of matrix $\boldsymbol{\beta}$ are distinct.
  \begin{IEEEeqnarray}{c}
  \boldsymbol{\beta}'=
  \left[\begin{array}{ccc}
    \beta_{1,1} & \ldots & \beta_{1,\Delta}   \\
    \vdots & \ddots & \vdots  \\
    \beta_{L ,1} & \ldots & \beta_{L ,\Delta}   \\
  \end{array}
  \right].
  \end{IEEEeqnarray}
  Then, the entries in columns $[(s-1)\Delta+1:s\Delta]$ of the matrix $\boldsymbol{\beta}$ are given by
  \begin{IEEEeqnarray}{rCl}
    \left[
  \begin{array}{ccc}\label{field:11}
    \beta_{1,(s-1)\Delta+1} & \ldots & \beta_{1,s\Delta}   \\
    \vdots & \ddots & \vdots  \\
    \beta_{L -s+1,(s-1)\Delta+1} & \ldots & \beta_{L -s+1,s\Delta}   \\
    \beta_{L -s+2,(s-1)\Delta+1} & \ldots &\beta_{L -s+1,s\Delta} \\
    \vdots & \ddots & \vdots  \\
    \beta_{L ,(s-1)\Delta+1} & \ldots & \beta_{L ,s\Delta}   \\
  \end{array}
  \right]=
  \left[
  \begin{array}{ccc}
    \beta_{s,1} & \ldots & \beta_{s,\Delta}  \\
    \vdots & \ddots & \vdots  \\
    \beta_{L ,1} & \ldots &\beta_{L ,\Delta} \\
    \beta_{1,1} & \ldots & \beta_{1,\Delta}   \\
    \vdots & \ddots & \vdots  \\
    \beta_{s-1,\Delta} & \ldots &\beta_{s-1,\Delta} \\
  \end{array}
  \right],\quad\forall\,s\in[2:S],
  \end{IEEEeqnarray}
  which is equivalent to performing upward circular shift of the matrix $\boldsymbol{\beta}'$ by $s-1$ positions.
\end{itemize}
  \item [Step 2.] Let $\beta_{1,K+1},\ldots,\beta_{1,K+X}$ be $X$ distinct elements from $\mathbb{F}_q\backslash\{\beta_{\ell,k}:\ell\in[L],k\in[K]\}$, i.e.,
  \begin{IEEEeqnarray}{c}
  \{\beta_{1,k}:k\in[K+1:K+X]\}\cap\{\beta_{\ell,k}:\ell\in[L],k\in[K]\}=\emptyset.
  \end{IEEEeqnarray}
  Then, set
  \begin{IEEEeqnarray}{c}
  (\beta_{\ell,K+1},\beta_{\ell,K+2},\ldots,\beta_{\ell,K+X})=(\beta_{1,K+1},\beta_{1,K+2},\ldots,\beta_{1,K+X}),\quad\forall\,\ell\in[2:L ].
  \end{IEEEeqnarray}

  \item [Step 3.]  Let
  \begin{IEEEeqnarray}{c}
  (\alpha_{1},\alpha_{2},\ldots,\alpha_{X})=(\beta_{1,K+1},\beta_{1,K+2},\ldots,\beta_{1,K+X})
  \end{IEEEeqnarray}
  and
  $\alpha_{X+1},\alpha_{X+2},\ldots,\alpha_{N}$ be another $N-X$ distinct elements from $\mathbb{F}_q$ such that
  \begin{IEEEeqnarray}{c}\label{field:1111}
  \{\alpha_{n}:n\in[X+1:N]\}\cap\{\beta_{\ell,k},\alpha_n:\ell\in[L],k\in[K],n\in[X]\}=\emptyset.
  \end{IEEEeqnarray}
\end{basedescript}

It is easy to prove that the constructed elements $\{\beta_{\ell,k},\alpha_n:\ell\in[L],k\in[K+X],n\in[N]\}$ in \eqref{field:1}-\eqref{field:1111} satisfy P1-P4 with
\begin{IEEEeqnarray}{c}\notag
|\{\beta_{\ell,k},\alpha_n:\ell\in[L],k\in[K+X],n\in[N]\}|=\max\{K, E\}+N.
\end{IEEEeqnarray}


Consequently, by \eqref{finite field size}, Lemma \ref{theorem:tield size} is proved.

\end{appendix}



\end{document}